\documentclass[a4paper,12pt]{article}
\usepackage[margin=1in]{geometry}
\usepackage{titlesec}
\usepackage{lipsum}
\usepackage[utf8]{inputenc}

\usepackage{booktabs} 
\usepackage{threeparttable} 

\usepackage{graphicx} 
\usepackage{caption} 
\usepackage{subcaption}
\usepackage{changepage}
\usepackage{adjustbox}
\usepackage{natbib}
\usepackage{amsfonts} 
\usepackage{amsmath}
\titleformat{\section}[block]{\normalfont\Large\bfseries}{\thesection}{1em}{}
\titleformat{\subsection}[block]{\normalfont\large\bfseries}{\thesubsection}{1em}{}
\usepackage{lscape} 

\usepackage{graphicx}
\usepackage{caption}
\usepackage{subcaption}
\usepackage{float} 

\makeatletter
\renewcommand\@biblabel[1]{}

\makeatother
 
\begin{document}

\title{Regime Changes and Real-Financial Cycles: Searching Minsky's Hypothesis in a Nonlinear Setting}
\author{
   Domenico Delli Gatti  \\ \smallskip
   \textit{CLE, Universit\`a Cattolica del Sacro Cuore, Milano} \\  
   Filippo Gusella \\ 
   \textit{CLE, Universit\`a Cattolica del Sacro Cuore, Milano}  \\ 
   \textit{Universit\`a degli Studi di Firenze} \\ \smallskip
   \textit{New York University in Florence} \\
   Giorgio Ricchiuti \\
   \textit{Universit\`a degli Studi di Firenze} \\
   \vspace{0.2em}
   \textit{CLE, Universit\`a Cattolica del Sacro Cuore, Milano}
}

\date{} 

\maketitle

\begin{abstract}
This paper investigates Minsky's cycles by extending the paper of \citet{stockhammer2019short} with a nonlinear model to capture possible local real-financial endogenous cycles. We trace nonlinear regime changes and check the presence of Minsky cycles from the 1970s to 2020 for the USA,  France, Germany, Canada, Australia, and the UK, linking the GDP with corporate debt, interest rate, and household debt. When considering corporate debt, the results reveal real-financial endogenous cycles in all countries, except Australia, and across all countries when interest rates are included. We find evidence for an interaction mechanism between household debt and GDP only for the USA and the UK. These findings underscore the importance of nonlinear regime transitions in empirically assessing Minsky's theory.
\end{abstract}
\bigskip
\bigskip

\bigskip
\bigskip

\textbf{Jel codes}: E32, G01
\bigskip
\bigskip

\newpage

\section{Introduction}
The role of the financial system lies at the heart of macroeconomic analysis.  Understanding how financial variables evolve and interact with real variables, driving economic growth, but also episodes of instability, has increasingly attracted the attention of economists. Among the theories that highlight this centrality, Hyman Minsky’s financial instability hypothesis (FIH) stands out as particularly influential \citep{minsky1978financial}. This theory emphasizes the crucial role that financial factors play in shaping the dynamics of business cycles: As real activity expands, it gradually creates a financially fragile environment, subsequently casting a negative influence on the real economy. This mechanism is at the core of Minsky’s hypothesis and provides a framework for understanding the real-financial interaction.

The literature on Minsky’s theory is largely dominated by theoretical contributions with limited empirical validation. From a theoretical point of view, the interaction mechanism à la Minsky has been analyzed in a wide and heterogeneous literature \citep[see, e.g.][for an overview]{nikolaidi2018minsky}. This body of work has been formalized through different theoretical frameworks, including nonlinear dynamic macroeconomic models, agent-based models, and stock-flow consistent models. Among them, some models emphasize the role of the interest rate in relation to the dynamics of corporate or household debt \citep[see, among others,][] {semmler1987macroeconomic,lima2007macrodynamics,fazzari2008cash,gatti2010financial,riccetti2013leveraged,riccetti2015agent,dafermos2018debt,reissl2020minsky,kohler2019exchange}, while another strand highlights the destabilizing dynamics of asset prices \citep[see, for example,][]{taylor1985minsky,gatti1990financial,ryoo2010long,chiarella2011financial,riccetti2016stock,alessia2021ir,gusella2025financial}. Despite these differences, the common idea is that financial fragility builds endogenously in periods of expansion, sowing the seeds for subsequent economic recession. 
                              
From an empirical perspective, only a limited number of studies have attempted to investigate the instability hypothesis. For instance, at the micro level, some works aim to discern hedge, speculative, and Ponzi states that characterize a firm's financial condition in various countries and economic sectors \citep{schroeder2009defining,nishi2012dynamic,mulligan2013sectoral,davis2019empirical}.\footnote{The financial instability hypothesis originally focused on firms’ investment and financing behavior. In Minsky’s framework, firms move from hedge positions - where cash flows are sufficient to meet debt obligations - to speculative positions - where refinancing becomes necessary - and finally to Ponzi positions, which rely on rising asset prices to service debt; this progressive shift for most firms makes the economy systemically fragile and ultimately leads to financial and economic crises.} 
Conversely, at the macro level, empirical literature can be broadly grouped into two main strands. A first group interprets macrofinancial fluctuations as the result of exogenous shocks hitting the financial sector, which subsequently propagate to the real economy \citep{kim2016macroeconomic,ma2016financial}. In contrast, a more recent contribution by \citet{stockhammer2019short}, and subsequent studies building on this approach \citep{kohler2022estimating,kohler2023flexible,stockhammer2023debt}, adopt an explicitly endogenous perspective, framing the interaction between the financial and real sides within a vector autoregressive (VAR) setting, which allows empirical assessment of the mathematical condition for the existence of endogenous Minskyan cycles. 

Despite the distinction between exogenous and endogenous mechanisms as the source of Minskyan cycles, both macro-empirical approaches remain grounded in a linear analytical structure. This hypothesis constrains the ability to capture nonlinear transitions and regime-dependent dynamics. In fact, the linear specification rules out the possibility that different macro-financial dynamics may emerge across distinct time periods, potentially overlooking cyclical patterns that are temporally concentrated. To address these limitations, we extend the paper by \citet{stockhammer2019short}, introducing a bivariate Markov-switching vector autoregressive model (MS-VAR) that describes the dynamic behavior of economic time series in the presence of possible nonlinear endogenous regime changes \citep{krolzig1997markov, hamilton2020time, hamilton2016macroeconomic}. In this framework, we distinguish between two regimes: one involves an interaction mechanism between the real and financial variables, allowing for a possible estimation of Minsky cycles. The other regime, in contrast, assumes no interaction between real and financial variables. This modeling approach allows us to capture the occurrence of real-financial interactions in line with Minsky’s theory within a regime-changing nonlinear context, providing valuable insight into its temporal and local dynamics. In doing so, we contribute to the literature by providing the first empirical implementation of a nonlinear estimation framework explicitly designed to identify Minsky cycles within macrofinancial dynamics. Furthermore, by applying filtering techniques to uncover state dynamics, we can trace regime changes and observe the manifestation of financial instability within specific years. Consequently, one structure, for example, the ``Minsky regime" (``No Minsky regime"), may dominate for a specific period until it is replaced by the ``No Minsky regime" (``Minsky regime") when the switching takes place. The time-varying filtering technique, applied across different countries and financial variables, offers a novel comparative perspective on how endogenous instability mechanisms emerge across various institutional and financial environments over different time periods.

Estimation is conducted using data from the USA, France, Germany, Canada, Australia, and the UK from the 1970s to 2020. We consider the gross domestic product as the real variable, while the non-financial corporate debt, the housing debt, and the short-term  interest rate are the financial variables. When focusing on corporate debt, the results indicate the presence of local Minsky cycles in all countries, with the sole exception of Australia. When short-term interest rates are incorporated into the analysis, evidence of such interaction mechanisms emerges consistently across all countries examined. In contrast, the relationship between household debt and GDP appears more limited. In this case, significant endogenous Minsky cycles are identified only for the United States and the United Kingdom, suggesting that the macroeconomic relevance of household debt may be more context-dependent and institutionally specific.

The remainder of the paper is structured as follows. Section 2 provides the mathematical framework and the empirical strategy for studying real-financial cycles in a nonlinear estimation context. Section 3 presents the data and the estimation results. Section 4 concludes. 
\newpage
\section{Methodology}

In this section, we introduce the proposed approach for the analysis of real-financial cycles. Based on \citet{stockhammer2019short}, the mathematical framework is presented in Section 2.1. Then, in Section 2.2, we extend the empirical estimation strategy from a linear to a nonlinear setting.

\subsection{Minsky cycles}

Minsky's theory suggests the interdependence between real and financial dynamics. The theory can be summarized in the famous sentence \emph{``stability is destabilizing"}. A period of economic expansion, characterized by an increase in GDP, entails an upward shift in debt and risk attitude. As financial fragility grows, a debt overhang or an increase in interest rates is reflected in a decline of the economic expansion \citep{stockhammer2019short}.

To formalize the cycle-generating interaction mechanism, the real variable ($y_t$) and the financial variable ($f_t$) are integrated into a simple first-order bivariate system of difference equations:

\begin{equation}\label{Eq.1}
\left[ {\begin{array}{*{20}{c}}
{{y_t}}\\
{{f_t}}
\end{array}} \right] = A{\left[ {\begin{array}{*{20}{c}}
{{y_{t - 1}}}\\
{{f_{t - 1}}}
\end{array}} \right]_,}\quad \quad with\quad \quad A = {\left[ {\begin{array}{*{20}{c}}
{{\alpha _1}}&{{\alpha _2}}\\
{{\beta _1}}&{{\beta _2}}
\end{array}} \right]_.}
\end{equation}

The dynamics of the system is given by the transition equation $A$, which describes the evolution of the real and financial variables. Eigenvalue analysis can be performed to investigate the conditions for potential oscillations in a two-dimensional discrete dynamical system. 

The eigenvalues $\lambda$ of the system satisfy the following characteristic equation:

\[\det \left( {A - \lambda I} \right) = 0,\]
i.e.:

\[\det \left[ {\begin{array}{*{20}{c}}
{{\alpha _1} - \lambda }&{{\alpha _2}}\\
{{\beta _1}}&{{\beta _2} - \lambda }
\end{array}} \right]=0_.\]

We obtain:

\[\left( {{\alpha _1} - \lambda } \right)\left( {{\beta _2} - \lambda } \right) - {\beta _1}{\alpha _2}=0,\]
from which:

\[{\lambda ^2} - \lambda Tr\left( A \right) - \det \left( A \right).\]

The roots assume the following form:

\[{\lambda _{1,2}} = \frac{{Tr\left( A \right) \pm \sqrt {Tr{{\left( A \right)}^2} - 4\det \left( A \right)} }}{2}.\]

The condition for oscillations is expressed in terms of the discriminant, which must be negative for the existence of complex eigenvalues. This criterion translates as follows:

\[\begin{array}{l}
\Delta  = Tr{\left( A \right)^2} - 4\det \left( A \right) < 0\\
\;\;\; \;= {\left( {{\alpha _1} + {\beta _2}} \right)^2} - 4\left( {{\alpha _1}{\beta _2} - {\alpha _2}{\beta _1}} \right)<0\\

\end{array}\]
from which the following condition must be satisfied:

\[{\left( {{\alpha _1} - {\beta _2}} \right)^2} + 4{\alpha _2}{\beta _1} < 0\]

From the previous equation, we can see that the necessary condition for oscillations is ${\alpha _2}{\beta _1}<0$. In other words, fluctuations in the system described by Eq. \ref{Eq.1} occur only if the interaction between the two variables is such that an increase in one leads to a rise in the second, which then pulls the first down.  In Minsky's vision of financial-real cycles, the usual assumption is that a rise in GDP has a positive effect on financial variables (${\beta _1}>0$), such as interest rates or debt, while an increase in financial variables has a negative effect on the real variable (${\alpha _2}<0$).

\subsection{The estimation strategy}

Building on the theoretical model presented in the previous section, we extend it to a nonlinear context using a Markov Switching Vector Autoregressive (MS-VAR) model. This framework allows the dynamics of the system to evolve across different regimes, governed by a latent state variable, $s_t$, which follows a first order discrete-time Markov process. The general form of the MS-VAR model is given by:

\[
\mathbf{y}_t = \mathbf{A}(s_t)\mathbf{y}_{t-1} + \mathbf{\epsilon}_{t},
\]
where $\mathbf{y}_t \in \mathbb{R}^n$  is the vector variables at time \( t \), $\mathbf{A}(s_t) \in \mathbb{R}^{n \times n}$ is the regime-dependent coefficient matrix and \( \mathbf{\epsilon}_t \sim N(0, \Sigma_{s_t}) \) is the vector error term with the variance-covariance matrix \( \Sigma_{s_t} \in \mathbb{R}^{n \times n} \).

With respect to our case, while the baseline model initially accounted for a single regime, it now differentiates between two distinct regimes (i.e., $n=2$): one representing potential real-financial interaction and another with independent real-financial dynamics, with the absence of cyclical interaction dynamics. In a two-regime MS-VAR model, the system is represented as:
\[
\mathbf{y}_t = 
\begin{cases}
\mathbf{A}_1 \mathbf{y}_{t-1} + \mathbf{\epsilon}_{t} & \text{if } s_t = 1 \ (\text{possible real-financial interaction}) \\
\mathbf{A}_2 \mathbf{y}_{t-1} + \mathbf{\epsilon}_{t} & \text{if } s_t = 2 \ (\text{no real-financial interaction})
\end{cases}
\]

Where: 

\[
\mathbf{y}_t = \begin{bmatrix} y_t \\ f_t \end{bmatrix}_.
\]

The coefficient matrix for regime 1 is:

\[
\mathbf{A}_1 = \begin{bmatrix}
\alpha_1 & \alpha_2 \\
\beta_1 & \beta_2
\end{bmatrix}_,
\]
with the following error term: 

\[
\mathbf{\epsilon}_{t} = \begin{bmatrix}
\varepsilon_t \\
\varphi_t
\end{bmatrix}, \quad \mathbf{\epsilon}_{t} \sim N(0, \Sigma_1).
\]

The coefficient matrix for regime 2 is:

\[
\mathbf{A}_2 = \begin{bmatrix}
\psi_1 & 0 \\
0 & \omega_2
\end{bmatrix}_,
\]
with the following error term: 

\[
\mathbf{\epsilon}_{t} = \begin{bmatrix}
\delta_t \\
\rho_t
\end{bmatrix}, \quad \mathbf{\epsilon}_{t} \sim N(0, \Sigma_2)_.
\]

In an extensive stochastic form, we obtain:
                
\[\left[ {\begin{array}{*{20}{c}}
{{y_t}}\\
{{f_t}}
\end{array}} \right] = \left\{ {\begin{array}{*{20}{l}}
{({s_t})\left[ {\begin{array}{*{20}{c}}
{{\alpha _1}}&{{\alpha _2}}\\
{{\beta _1}}&{{\beta _2}}
\end{array}} \right]\left[ {\begin{array}{*{20}{c}}
{{y_{t - 1}}}\\
{{f_{t - 1}}}
\end{array}} \right] + \left[ {\begin{array}{*{20}{c}}
{{\varepsilon _t}}\\
{{\varphi _t}}
\end{array}} \right]{\kern 1pt} \quad \text{(possible\quad real-financial\quad cycle)\quad} }\\
\begin{array}{l}
\\
({s_t})\left[ {\begin{array}{*{20}{c}}
{{\psi _1}}&0\\
0&{{\omega _2}}
\end{array}} \right]\left[ {\begin{array}{*{20}{c}}
{{y_{t - 1}}}\\
{{f_{t - 1}}}
\end{array}} \right] + \left[ {\begin{array}{*{20}{c}}
{{\delta _t}}\\
{{\rho _t}}
\end{array}} \right]\quad \text{(no\quad real-financial\quad cycle)}
\end{array}
\end{array}} \right.\]

In the first regime, with coefficients \(\alpha_1\), \(\alpha_2\), \(\beta_1\), and \(\beta_2\),  if ${\alpha _2}{\beta _1}<0$ there is a local cyclical interaction between real and financial variables with a feedback loop between financial conditions and real economic outcomes. If ${\beta _1}>0$ and ${\alpha _2}<0$ these cycles are Mynskian cycles. In contrast, the second regime, represented only by the coefficients in the main diagonal \(\psi_1\) and \(\omega_2\), captures periods when the real and financial variables follow independent paths, without cyclical interaction. Error terms ($\varepsilon _t$, $\varphi _t$, $\delta _t$ and $\rho _t$) are zero mean withe-noise processes with variance collected in the variance-covariance matrices $\Sigma_1$ and $\Sigma_2$. 

${s_t}$ is the latent state-space discrete-time Markov chain representing the switching mechanism among the two regimes (or states). The state variable $s_t$ follows the first regime (regime $1$)  when a possible real financial interaction is detected from the data and the second regime (regime 2) when the two variables follow an independent path.

There are four kinds of possible transitions between the two states:
\begin{itemize}
\item From  state 1 to state 1:  with probability ${p_{11}} = P\left( {{s_t} = 1\left| {{s_{t - 1}} = 1} \right.} \right)$\footnote{This is read as the probability that the system is in regime 1  at time
$t$, given that it was in the same regime at the previous time $(t-1)$.}
\item From  state 1 to state 2:  with probability ${p_{12}} = P\left( {{s_t} = 2\left| {{s_{t - 1}} = 1} \right.} \right)$
\item From  state 2 to state 1:  with probability ${p_{21}} = P\left( {{s_t} = 1\left| {{s_{t - 1}} = 2} \right.} \right)$
\item From  state 2 to state 2:  with probability ${p_{22}} = P\left( {{s_t} = 2\left| {{s_{t - 1}} = 2} \right.} \right)$
    
\end{itemize}
with:

\[{p_{11}} + {p_{12}} = 1\quad and\quad {p_{21}} + {p_{22}} = 1_.\]

In this way, ${s_t}$ depends on ${s_{t-1}}$ according to the following state transition matrix, which illustrates the probability of switching between these regimes over time.      

\[\left[ {\begin{array}{*{20}{c}}
{{p_{11}}}&{{p_{12}}}\\
{{p_{21}}}&{{p_{22}}}
\end{array}} \right]_.\]

\section{Dataset and estimation results}

We use data spanning from the 1970s to 2020 for six OECD countries: the United States, the United Kingdom, France, Germany, Canada, and Australia. The real variable $y$ is proxied by seasonally adjusted real gross domestic product (GDP), obtained from the OECD Statistics and transformed into logarithmic levels. As financial variables $f$, we consider nonfinancial corporate debt (NFCD), housing debt (HD), and the short-term nominal interest rate (STIR). Data on NFCD and HD are sourced from the Bank for International Settlements (BIS) Data Portal, while STIR is obtained from the OECD Statistics.\footnote{For the UK, the nominal interest rate is recovered from Federal Reserve Economic Data (FRED).}

All the data are at yearly frequency, a choice motivated by several considerations. First, from a theoretical perspective, Minsky’s framework addresses medium- to long-term dynamics, which are more appropriately analyzed with lower-frequency data.\footnote{This choice also aligns with the original analysis by \citet{stockhammer2019short}, which employs annual data.} Second, from an empirical standpoint, yearly data mitigates serial correlation in the errors. In contrast, higher-frequency data would exacerbate serial correlation, requiring the inclusion of lag operators. This adjustment would increase the dimensionality of the system and preclude the recovery of a simple mathematical condition for cyclical dynamics. Furthermore, within a nonlinear framework, additional lag operators would place excessive demands on the sample. To ensure tractability and reliable estimation, we therefore focus on a selective estimation  of the most essential parameters \citep{hamilton2016macroeconomic}. Finally, we address the small-sample issue by conducting 100 Monte Carlo simulations to verify the robustness of our results.

Once data are collected,  we focus on cyclical phenomena by first extracting cycles from the time series using the Hodrick-Prescott filter, setting the smoothing parameter suggested for the yearly frequency \citep{hodrick1997postwar}. This also allows us to transform our series into a stationary series, thereby maximizing the likelihood function. Concerning this last point, the Markov switching estimation is carried out using the expectation-maximization algorithm \citep{krolzig1997markov, hamilton2020time, hamilton2016macroeconomic}. The iterative EM process consists of an expectation (E) step, where expected latent variable values are calculated based on current parameters, followed by a maximization (M) step, which identifies parameter values that enhance the expected log-likelihood from the E step. The parameters are used to recover the latent states' distribution in the next E step.

\subsection{GDP/NFCD interaction}

Table \ref{GDP/NFCD} presents the estimation results when GDP and NFCD are included in the system.\footnote{As previously specified, the cyclical component of the two series is obtained by implementing the HP filter, which renders the series stationary and allows us to maximize the log-likelihood function. The results in Appendix A confirm the stationarity of the series for all the cases considered.} The table is structured to display the coefficients for the two regimes with the associated transition matrix.

\begin{table}[ht]
\begin{threeparttable}
\centering
\caption{Estimation Results for GDP/NFCD}\label{GDP/NFCD}
\medskip

\begin{tabular}{cccc}

\hline\hline\\[-1.8ex]       
GDP/NFCD & Regime 1 & Regime 2 & Transition Matrix \\
\hline\hline\\[-1.8ex] 
USA & \begin{tabular}[c]{@{}cc@{}} $0.8215^{***}$& $-0.1066^{***}$\\ (0.0678) & (0.0255) \\ $1.6348^{***}$& $0.4027^{***}$ \\ (0.0884) & (0.0333) \end{tabular} & \begin{tabular}[c]{@{}cc@{}} $0.5389^{***}$ &  \\ (0.1330) &  \\  & $0.9884^{***}$  \\  & (0.0669) \end{tabular} & \begin{tabular}[c]{@{}cc@{}}  &  \\ 0.837 & 0.163  \\ 0.132  & 0.868  \\ & \end{tabular} \\
\hline
& & & \\ 
\hline
UK & \begin{tabular}[c]{@{}cc@{}} $0.8541^{***}$& $-0.0924^{***}$\\ (0.0578) & (0.0170) \\ $1.4605^{***}$& $0.7026^{***}$ \\ (0.2292) & (0.0676) \end{tabular} & \begin{tabular}[c]{@{}cc@{}} $0.04429^{}$ &  \\ (0.1180) &  \\  & $0.5760^{***}$  \\  & (0.0797) \end{tabular} & \begin{tabular}[c]{@{}cc@{}}  &  \\ 0.900 & 0.100  \\ 0.310  & 0.690   \\ & \end{tabular} \\
\hline
& & & \\ 
\hline
France & \begin{tabular}[c]{@{}cc@{}} $0.9783^{***}$& $-0.1284^{***}$\\ (0.1361) & (0.0418) \\ $1.6935^{***}$& $0.7408^{***}$ \\ (0.1516) & (0.0466) \end{tabular} & \begin{tabular}[c]{@{}cc@{}} $0.3312^{***}$ &  \\ (0.0539) &  \\  & $0.8120^{***}$  \\  & (0.0953) \end{tabular} & \begin{tabular}[c]{@{}cc@{}}  &  \\ 0.641 & 0.359  \\ 0.388  & 0.612    \\ & \end{tabular} \\
\hline
& & & \\ 
\hline              
Germany & \begin{tabular}[c]{@{}cc@{}} $0.5682^{***}$& $-0.1550^{***}$\\ (0.1160) & (0.0481) \\ $0.59835^{***}$& $0.8093^{***}$ \\ (0.1373) & (0.0569) \end{tabular} & \begin{tabular}[c]{@{}cc@{}} $0.6848^{***}$ &  \\ (0.0719) &  \\  & $0.5400^{***}$  \\  & (0.1249) \end{tabular} & \begin{tabular}[c]{@{}cc@{}}  &  \\ 0.945 & 0.055   \\ 0.150   & 0.850    \\ & \end{tabular} \\
\hline                
& & & \\ 
\hline
Canada & \begin{tabular}[c]{@{}cc@{}} $0.0246^{}$& $-0.30430^{***}$\\ (0.0768) & (0.0309) \\ $1.525^{***}$& $0.6036^{***}$ \\ (0.2236) & (0.0899) \end{tabular} & \begin{tabular}[c]{@{}cc@{}} $0.5951^{***}$ &  \\ (0.1281) &  \\  & $0.4882^{***}$  \\  & (0.1119) \end{tabular} & \begin{tabular}[c]{@{}cc@{}}  &  \\ 0.630 & 0.370   \\ 0.311  & 0.689   \\ & \end{tabular} \\
\hline      
& & & \\ 
\hline
Australia & \begin{tabular}[c]{@{}cc@{}} $0.83897^{***}$& $0.0110^{}$\\ (0.199) & (0.0353) \\ $3.2656^{***}$& $0.7459^{***}$ \\ (0.5873) & (0.1040) \end{tabular} & \begin{tabular}[c]{@{}cc@{}} $-0.0888^{*}$ &  \\ (0.0491) &  \\  & $0.7009^{***}$  \\  & (0.0578) \end{tabular} & \begin{tabular}[c]{@{}cc@{}}  &  \\ 0.507 & 0.493    \\ 0.695  & 0.305   \\ & \end{tabular} \\
\hline    
& & & \\ 
\hline
\end{tabular}
\footnotesize
 \begin{tablenotes}

            \item $^{***}$, $^{**}$, $^{*}$ are significance level at 1\%, 5\% and 10\%.
            \item Standard errors are in parenthesis.
            \item In regime one, regime two and the transition matrix, the reported values follow the positions of the parameters in section 2.1.   
            
        \end{tablenotes}
    \end{threeparttable}
\end{table}

\clearpage

For all countries, except Australia, the mathematical condition to obtain complex eigenvalues is respected [${\left( {{\alpha _1} - {\beta _2}} \right)^2} + 4{\alpha _2}{\beta _1}<0$]. The necessary condition (${\alpha _2}{\beta _1}<0$) is satisfied with cyclical parameters significant at a one percent statistical level. For Australia, signs are not respected, and parameters are not significant. Moreover, for the USA, France, Germany, Canada and the UK, the signs of cyclical coefficients ($\alpha_2<0$ and $\beta_1>0$) in regime one lead to the generation of endogenous Minsky cycles, a cyclical mechanism where an increase in the real variable leads to a subsequent rise in the financial variable, which in turn results in a decline in the real component. Contrary to regime one, regime two involves the absence of real-financial interaction between the real and the financial variable. The coefficients of the lagged values of the variables are significant for most countries, with the exception of Australia and the UK. For both regimes, the diagnostic check for autocorrelation is performed on the error terms following \citet{krolzig1997markov}. Serial correlation tests (refer to Appendix B) indicate that MS-VAR models do not exhibit autocorrelation for all countries considered.

To verify the robustness of our estimates, we conduct a Monte Carlo simulation study. To implement it, we generate $n = 100$ sample paths of observations from the estimated model by randomly generating state disturbances from the standard normal distribution and incorporating them into the nonlinear regime-switching model. We repeat the estimation process 100 times. Once we obtain the results, we calculate the mean value of the parameters to determine the presence of Minskyan fluctuations. Consistent with previous findings, for the USA, France, Germany, Canada, and the UK, the results confirm the presence of endogenous real-financial interaction à la Minsky, alternating with a regime of no interaction between the real and financial variables (see Appendix C).

Shifting our focus toward the transition matrices, they reveal interesting insights into the persistence of the two regimes. From Table \ref{GDP/NFCD}, the USA, Germany and the UK show the highest probabilities of remaining within the same regime (e.g., for regime 1, $p_{11}=0.837$ in the USA, $p_{11}=0.945$ for Germany, and $p_{11}=0.900$ in the UK), suggesting a stable regime structure over time. In contrast, France and Canada exhibit more frequent state transitions, indicating a greater probability of shifting between the two different regimes.   

\begin{figure}[htp]
    \centering
    \includegraphics[width=0.8\linewidth]{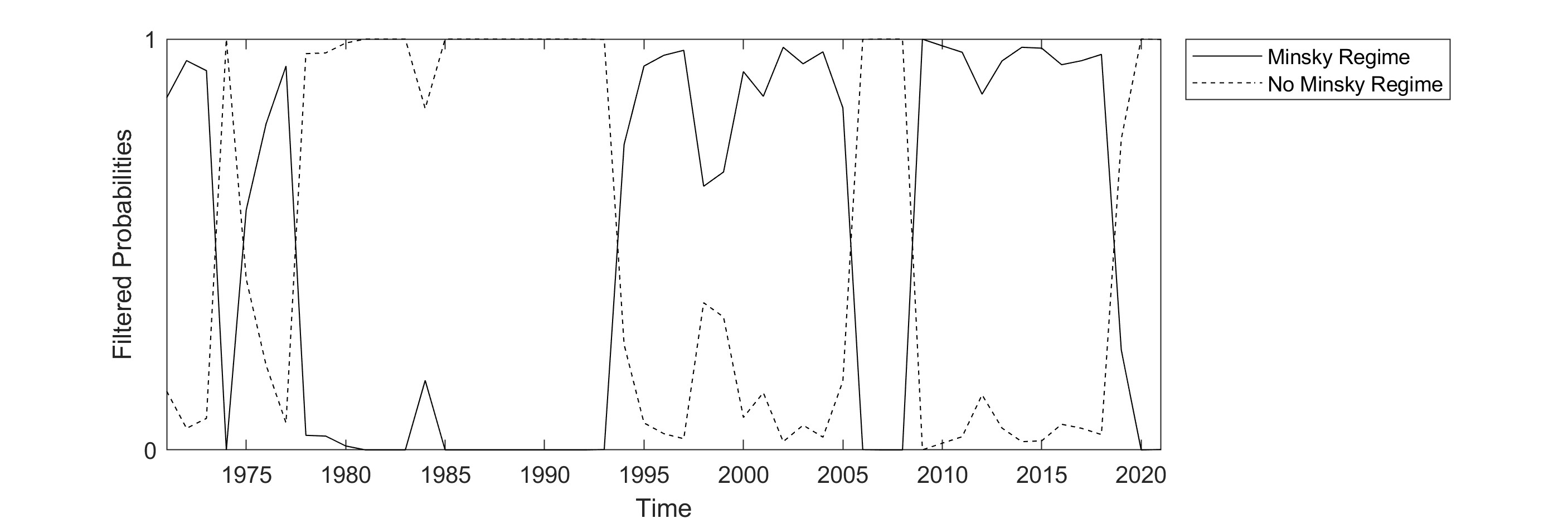}
    \caption{Filtered probability of the two regimes in the USA.}\label{filtered_NFCD1}
\end{figure}

\begin{figure}[htp]
    \centering
    \includegraphics[width=0.8\linewidth]{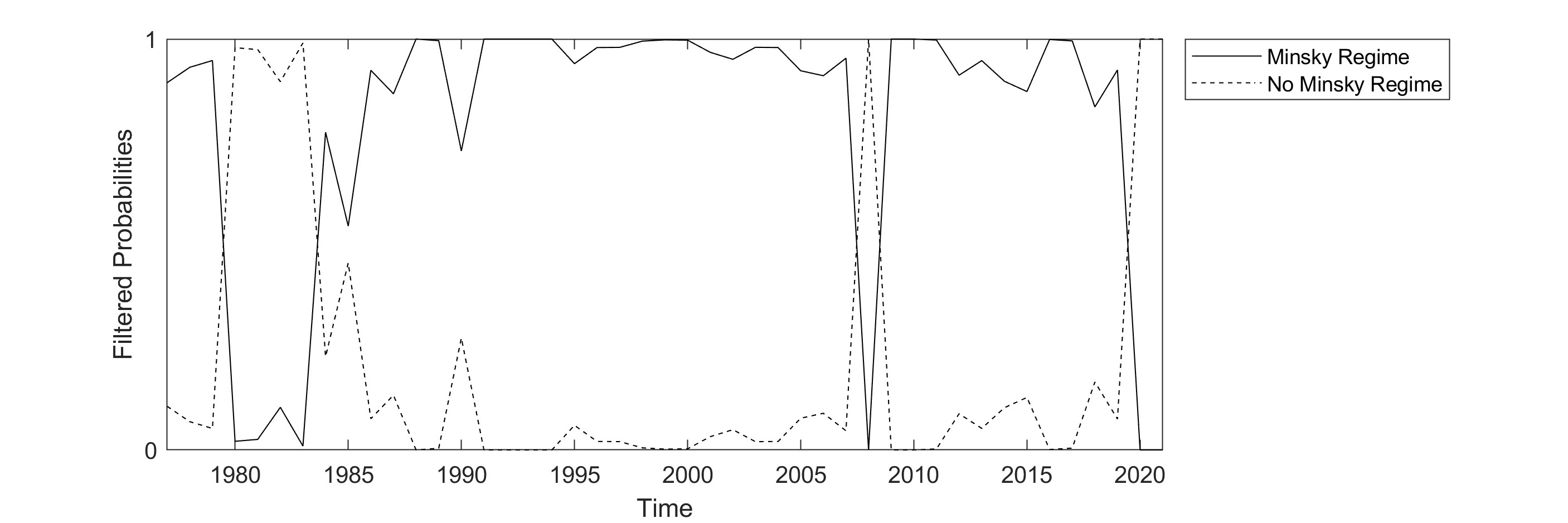}
    \caption{Filtered probability of the two regimes in the UK.}\label{filtered_NFCD2}
\end{figure}

\begin{figure}[htp]
    \centering
    \includegraphics[width=0.8\linewidth]{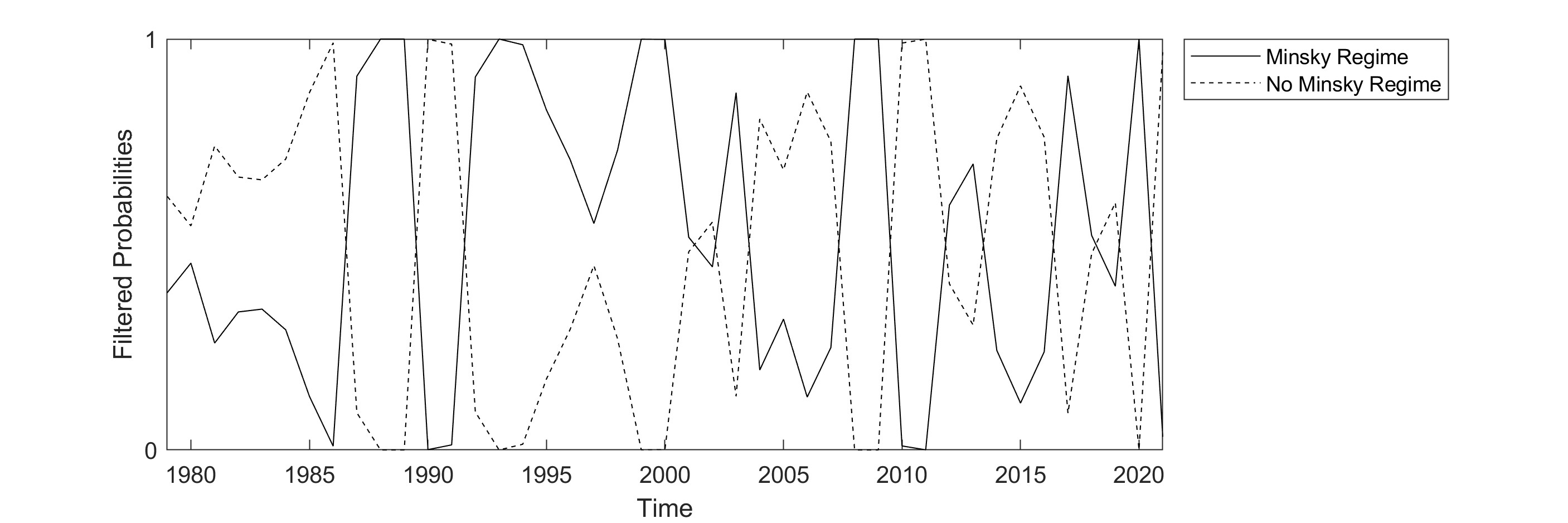}
    \caption{Filtered probability of the two regimes in France.}\label{filtered_NFCD3}
\end{figure}

\begin{figure}[htp]
    \centering
    \includegraphics[width=0.8\linewidth]{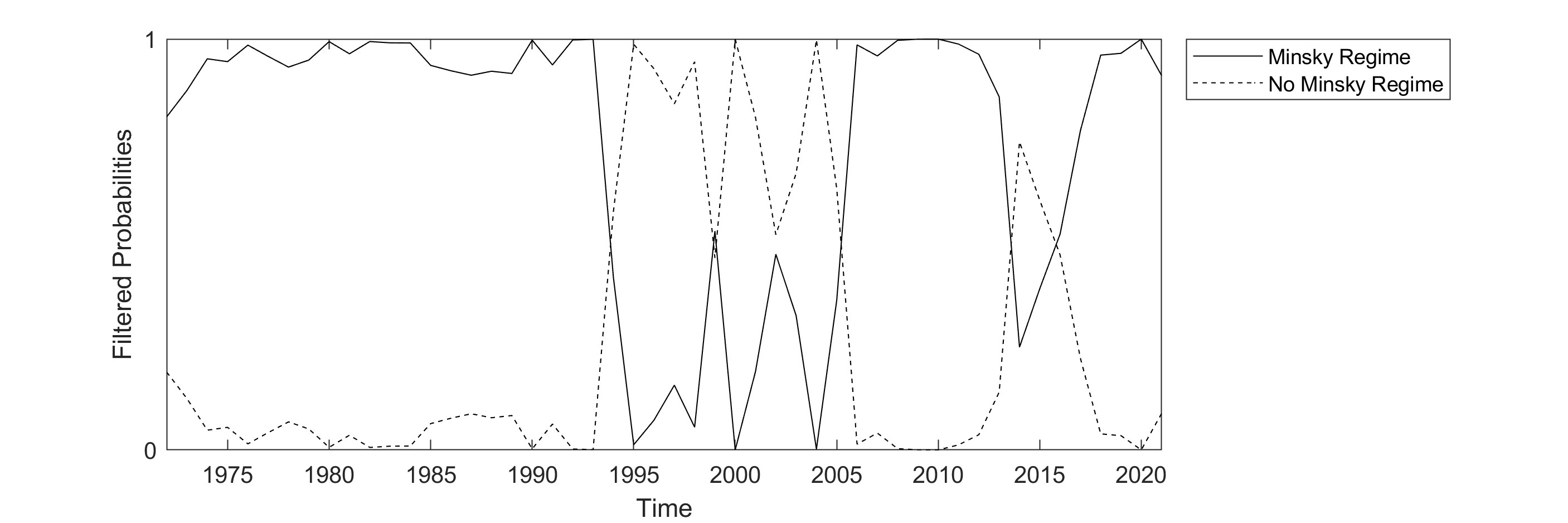}
    \caption{Filtered probability of the two regimes in Germany.}\label{filtered_NFCD4}
\end{figure}

\begin{figure}[htp]
    \centering
    \includegraphics[width=0.8\linewidth]{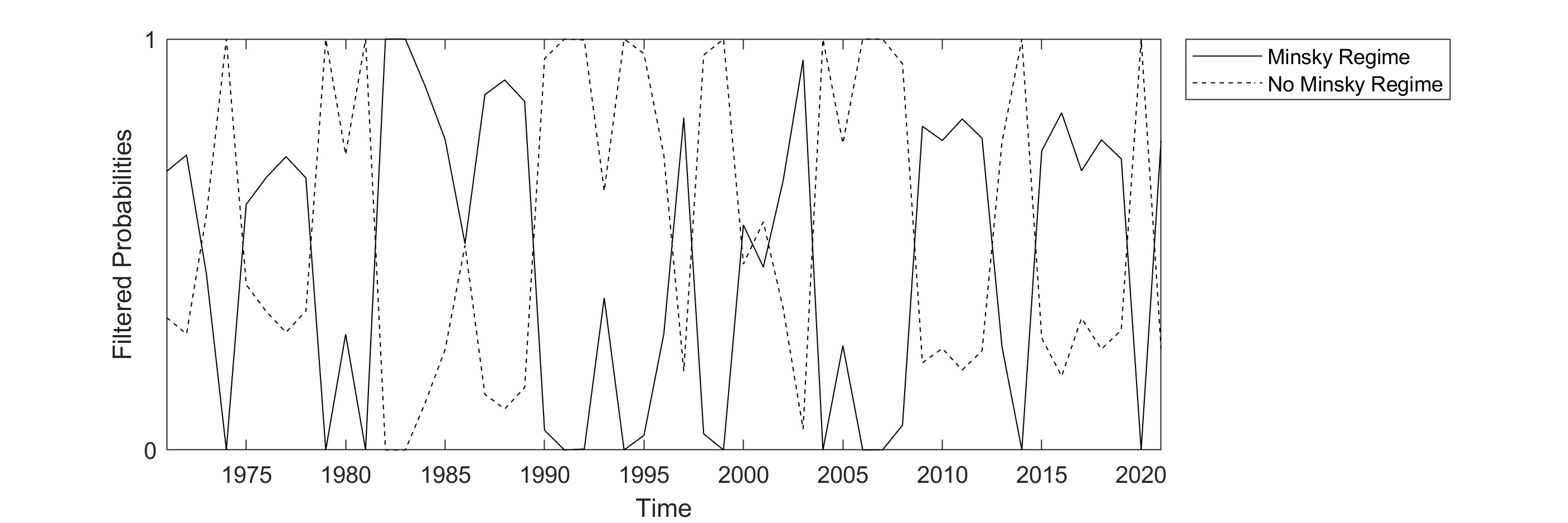}
    \caption{Filtered probability of the two regimes in Canada.}\label{filtered_NFCD5}
\end{figure}

Figures \ref{filtered_NFCD1}, \ref{filtered_NFCD2}, \ref{filtered_NFCD3}, \ref{filtered_NFCD4}, \ref{filtered_NFCD5} show the filtered probabilities of two regimes across the USA, the UK, France, Germany, and Canada from the 1970s to 2020, with the solid line representing the ``Minsky Regime" and the dashed line representing the ``No Minsky Regime". For the USA, from the 1970s to the early 1980s, the probability of being in the ``Minsky Regime" was relatively high. In contrast, from the mid-1980s to the early 1990s, the probabilities shifted significantly towards the ``No Minsky Regime", suggesting a period of independence between real and financial variables. This pattern changed in the mid-1990s with a clear return to the ``Minsky Regime". This trend culminated in the explosion of the global financial crisis of 2007-2008, with a temporary increase in the ``No Minsky Regime" probability in the early 2010s. From 2010 to 2020, the graph shows alternating probabilities with a general trend toward the ``Minsky Regime." A similar pattern can be observed for the UK. The only difference is the starting period of the real/financial interaction à la Minsky. In fact, the UK experienced an increasing probability of the ``Minsky Regime" starting from the middle of the 1980s. As in the USA, this regime has been persistent around the dot-com crisis and during the global financial crisis. 

Figs. \ref{filtered_NFCD3} and \ref{filtered_NFCD5} for France and Canada indicate more volatile patterns compared to the USA and the UK. These fluctuations reflect a greater sensitivity to frequent transitions between the two regimes. Similar to the USA and the UK, France also shows high probabilities for the ``Minsky Regime" during the early 2000s and pre-2007/2008. For Germany, the ``Minsky Regime" dominated for an extended period from the late 1970s through the mid-1990s. Post-1995, there is a noticeable shift towards the ``No Minsky Regime",  interrupted around the global financial crisis.

Overall, the results highlight a time-dependent Minsky-type cyclical relationship between GDP and NFCD in most countries. From the filtered probabilities, this regime dominates essentially during the 1970s and in periods that culminated with the dot-com crisis and the global financial crisis.  Transition matrices further reveal heterogeneity across countries, with the UK, Germany, and the USA showing high regime persistence, while France and Canada display more frequent shifts.
\bigskip
\bigskip
\bigskip

\bigskip
\bigskip
\bigskip

\clearpage

\bigskip
\bigskip
\bigskip

\bigskip
\bigskip
\bigskip

\clearpage

\subsection{GDP/HD interaction}

Table \ref{GDP/HD} presents the estimation results from the MS-VAR model that examines the relationship between GDP and HD across the countries considered. 

\begin{figure}[htp]
    \centering
    \includegraphics[width=0.8\linewidth]{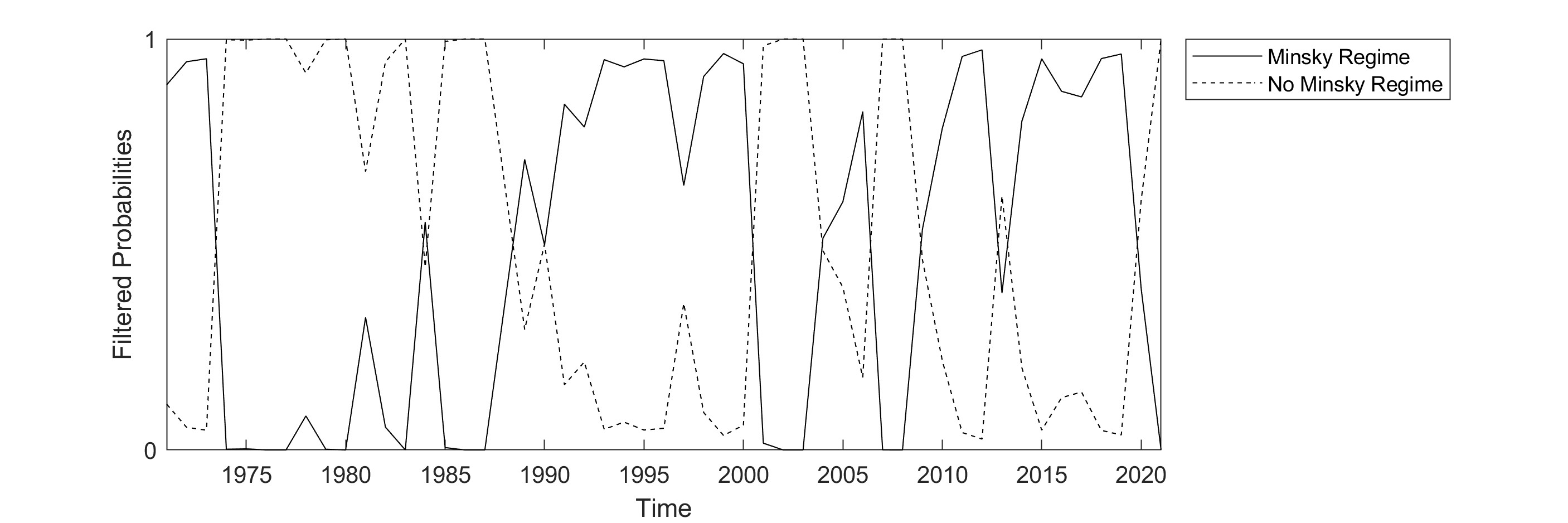}
    \caption{Filtered probability of the two regimes in the USA.}\label{filtered_HD1}
\end{figure}

\begin{figure}[htp]
    \centering
    \includegraphics[width=0.8\linewidth]{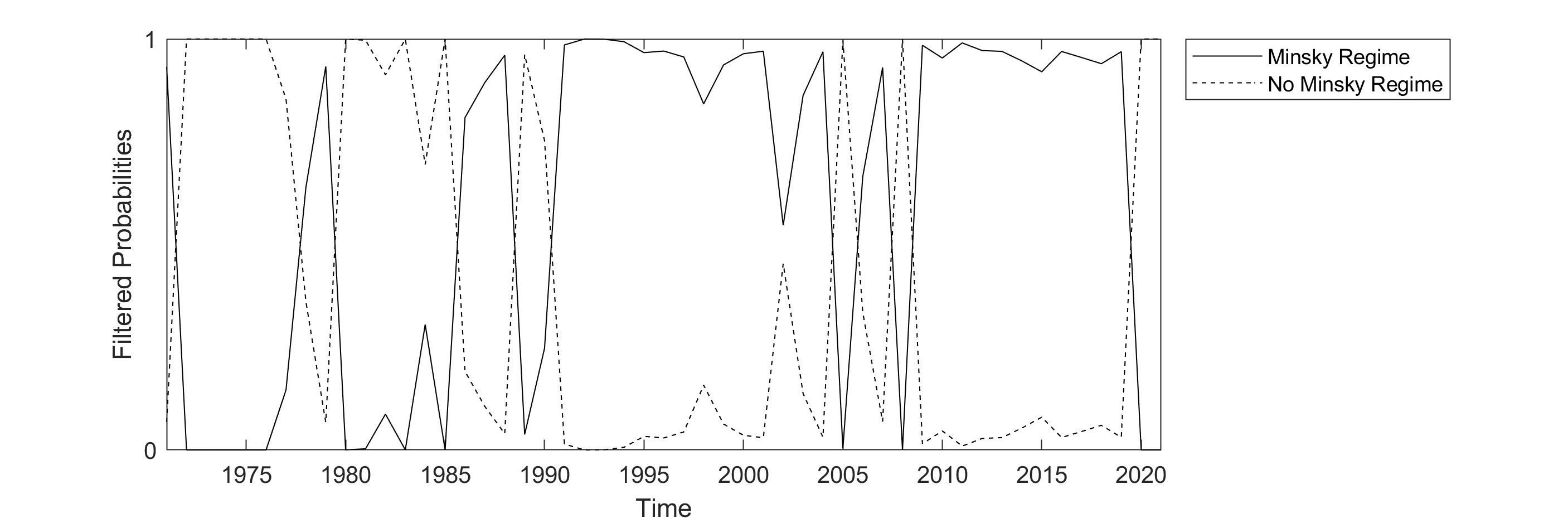}
    \caption{Filtered probability of the two regimes in the UK.}\label{filtered_HD2}
\end{figure}

In the USA, the cyclical coefficients for regime 1 are $\beta_1=0.7348$  and $\alpha_2=-0.1123$, significant at levels 1\% and 5\%, respectively, and with signs that reflect the generation of endogenous Minskyan cycles. This regime is persistent, with $p_{11}=0.824$, suggesting that once the system enters this state, it will likely remain there for an extended period. In particular, this regime emerged strongly during the 1990s, with peaks during the global financial crisis (Fig. \ref{filtered_HD1}). Similarly, the UK exhibits significant coefficients in regime 1: $\beta_1= 0.7925$, and $\alpha_2=-0.1763$  at the 1\% statistical level. As for the USA, the transition probability for regime 1 is $p_{11}=0.802$, indicating high persistence, especially for the 90s through the subsequent period (Fig. \ref{filtered_HD2}).

For Germany and Australia, regime one does not display the mathematical conditions to obtain endogenous cycles. In Germany,  $\beta_1= -0.0433$   and $\alpha_2= -0.2991$, while for Australia, they are $\beta_1= 0.4722$  and $\alpha_2=  0.2357$. In other words, the necessary condition ($\alpha \beta_1<0$) is not satisfied.  For France, although the mathematical conditions are satisfied, the coefficients are not statistically significant. Finally, for Canada, $\alpha_2=-0.21825$ and $\beta_1=0.18783$ respect the necessary condition to obtain cyclical conditions and  are statistically significant at one percent and five percent statistical levels, respectively, but the magnitudes are too small to sustain cyclical dynamics. 

As in the case with nonfinancial corporate debt, we tested the robustness of the results. Monte Carlo simulation results, summarized in Appendix C, provide further empirical evidence supporting the existence of distinct regimes with the existence of local endogenous Minskyan cycles for the UK and the USA, but not for the other countries considered. 

In summary, the introduction of household debt into the analysis indicates that Minsky's hypothesis is empirically found in the USA and the UK. Specifically, in regime 1, which persisted during the global financial crisis, the cyclical parameters are statistically significant and exhibit signs consistent with Minskyan cycle generation. In contrast, other countries show no evidence of Minsky cycles associated with household debt. For example, regime one does not meet the mathematical condition for endogenous cycles in Germany and Australia, parameters lack statistical significance in France, and the magnitudes are insufficient to generate cyclical phenomena in Canada. These findings highlight the unique economic dynamics in the USA and UK, where household debt played a critical role in generating Minskyan cycles, unlike in the other countries studied.

\begin{table}[ht]
\begin{threeparttable}
\centering
\caption{Estimation Results for GDP/HD}\label{GDP/HD}
\medskip

\begin{tabular}{cccc}

\hline\hline\\[-1.8ex]       
GDP/HD & Regime 1 & Regime 2 & Transition Matrix \\
\hline\hline\\[-1.8ex] 
USA & \begin{tabular}[c]{@{}cc@{}} $0.6615^{***}$& $-0.1123^{**}$\\ (0.1050) & (0.0437) \\ $0.7348^{***}$& $0.6325^{***}$ \\ 0.0802) & (0.0334) \end{tabular} & \begin{tabular}[c]{@{}cc@{}} $0.2878^{***}$ &  \\ (0.0994) &  \\  & $0.8359^{***}$  \\  & (0.0939) \end{tabular} & \begin{tabular}[c]{@{}cc@{}}  &  \\ 0.824 & 0.176 \\ 0.157  & 0.843 \\ & \end{tabular} \\
\hline
& & & \\ 
\hline
UK & \begin{tabular}[c]{@{}cc@{}} $0.7595^{***}$& $-0.1763^{***}$\\ (0.0756) & (0.0378) \\ $0.7925^{***}$& $0.7203^{***}$ \\ (0.0752) & (0.0376) \end{tabular} & \begin{tabular}[c]{@{}cc@{}} $-0.0203^{}$ &  \\ (0.1474) &  \\  & $0.1867^{}$  \\  & (0.1386) \end{tabular} & \begin{tabular}[c]{@{}cc@{}}  &  \\ 0.802 & 0.198   \\ 0.292  & 0.708  \\ & \end{tabular} \\
\hline
& & & \\ 
\hline
France & \begin{tabular}[c]{@{}cc@{}} $0.3992^{***}$& $-0.1136^{}$\\ (0.1196) & (0.1093) \\ $0.1560^{}$& $0.6310^{***}$ \\ (0.1090) & (0.0996) \end{tabular} & \begin{tabular}[c]{@{}cc@{}} $1.2871^{***}$ &  \\ (0.0496) &  \\  & $0.9872^{***}$  \\  & (0.0746) \end{tabular} & \begin{tabular}[c]{@{}cc@{}}  &  \\ 0.497 & 0.503  \\ 0.945  & 0.055    \\ & \end{tabular} \\
\hline
& & & \\ 
\hline              
Germany & \begin{tabular}[c]{@{}cc@{}} $1.0552^{***}$& $-0.2991^{***}$\\ (0.1465) & (0.0972) \\ $-0.0433^{***}$& $0.7763^{***}$ \\ (0.1600) & (0.1061) \end{tabular} & \begin{tabular}[c]{@{}cc@{}} $-0.0420^{***}$ &  \\ (0.0537) &  \\  & $0.8091^{***}$  \\  & (0.0415) \end{tabular} & \begin{tabular}[c]{@{}cc@{}}  &  \\ 0.331 & 0.669  \\ 1.000   & 0.000    \\ & \end{tabular} \\
\hline                
& & & \\ 
\hline
Canada & \begin{tabular}[c]{@{}cc@{}} $0.3706^{***}$& $-0.2182^{***}$\\ (0.1111) & (0.0676) \\ $0.1878^{**}$& $0.7923^{***}$ \\ (0.0957) & (0.0582) \end{tabular} & \begin{tabular}[c]{@{}cc@{}} $0.1344^{}$ &  \\ (0.0973) &  \\  & $0.8100^{***}$  \\  & (0.0602) \end{tabular} & \begin{tabular}[c]{@{}cc@{}}  &  \\ 0.937 & 0.063   \\ 0.086  & 0.914  \\ & \end{tabular} \\
\hline      
& & & \\ 
\hline
Australia & \begin{tabular}[c]{@{}cc@{}} $0.1638^{***}$& $0.2357^{***}$\\ (0.0499) & (0.0352) \\ $0.4722 ^{***}$& $0.6692^{***}$ \\ (0.0626) & (0.0442) \end{tabular} & \begin{tabular}[c]{@{}cc@{}} $0.6194^{***}$ &  \\ (0.1266) &  \\  & $0.4327^{***}$  \\  & (0.1192) \end{tabular} & \begin{tabular}[c]{@{}cc@{}}  &  \\ 0.495 & 0.505   \\ 0.253  & 0.747   \\ & \end{tabular} \\
\hline    
& & & \\ 
\hline
\end{tabular}
\footnotesize
 \begin{tablenotes}

            \item $^{***}$, $^{**}$, $^{*}$ are significance level at 1\%, 5\% and 10\%.
            \item Standard errors are in parenthesis.
            \item In regime one, regime two and the transition matrix, the reported values follow the positions of the parameters in section 2.1.
            
        \end{tablenotes}
    \end{threeparttable}

\end{table}

\clearpage

\subsection{GDP/STIR interaction}

Table \ref{GDP/STIR} presents the results of the MS-VAR model, which examines the interaction between GDP and the short-term interest rate. In regime 1, for all countries, the cyclical condition is respected (${\left( {{\alpha _1} - {\beta _2}} \right)^2} + 4{\alpha _2}{\beta _1}<0$). The two parameters of interest ($\alpha_2 $ and $\beta_1)$ exhibit signs of generating Minsky cycles ($\alpha_2<0$ and $\beta_1>0 $), indicating that an increase in the real variable leads to a subsequent increase in the interest rate, which eventually constrains GDP growth.

The transition matrices in Table  reveal a high probability of remaining in the two regimes for the USA, France, Australia and the UK, suggesting persistent Minsky cycles but also persistent independent processes. Canada and Germany show a more frequent transition between the two regimes, reflecting frequent periods of both real-financial interaction and autonomous processes. 

Similarly to our analysis with corporate and household debt, we extend our analysis with 100 Monte Carlo simulations. A summary of the results is presented in Appendix C. Overall, the estimation results provide empirical evidence supporting the presence of endogenous real-financial cycles in regime 1 for all the countries considered.

\begin{figure}[htp]
    \centering
    \includegraphics[width=0.8\linewidth]{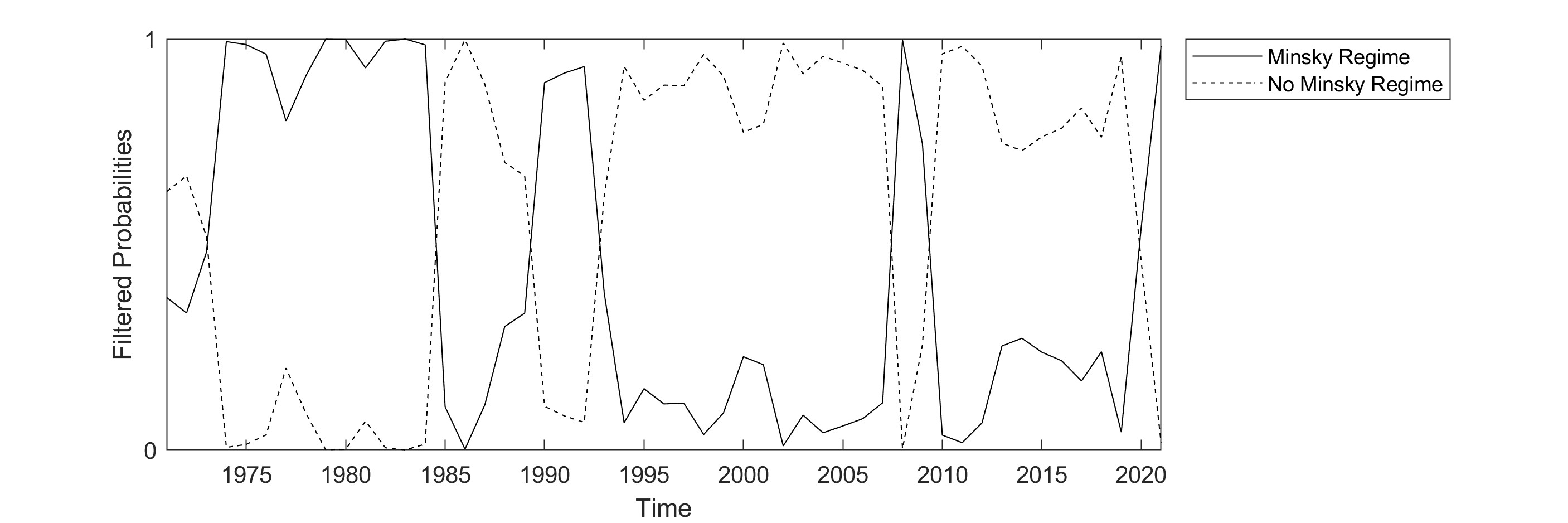}
    \caption{Filtered probability of the two regimes in the USA.}\label{filtered_Interest1}
\end{figure}

We now pass to analyze the filtered probabilities of the system being in two distinct regimes. For the USA (Fig. \ref{filtered_Interest1}), from the early 1970s to the mid-1980s, the ``Minsky Regime" appears to dominate, indicating frequent or prolonged periods of real-financial interactions. After the 1985, there are noticeable shifts between the two regimes, with some periods dominated by the ``No Minsky Regime", such as the mid-1990s and post-2010, while others,  in the corresponding of 2007/2008, reflect a resurgence of the ``Minsky Regime". 

\begin{figure}[htp]
    \centering
    \includegraphics[width=0.8\linewidth]{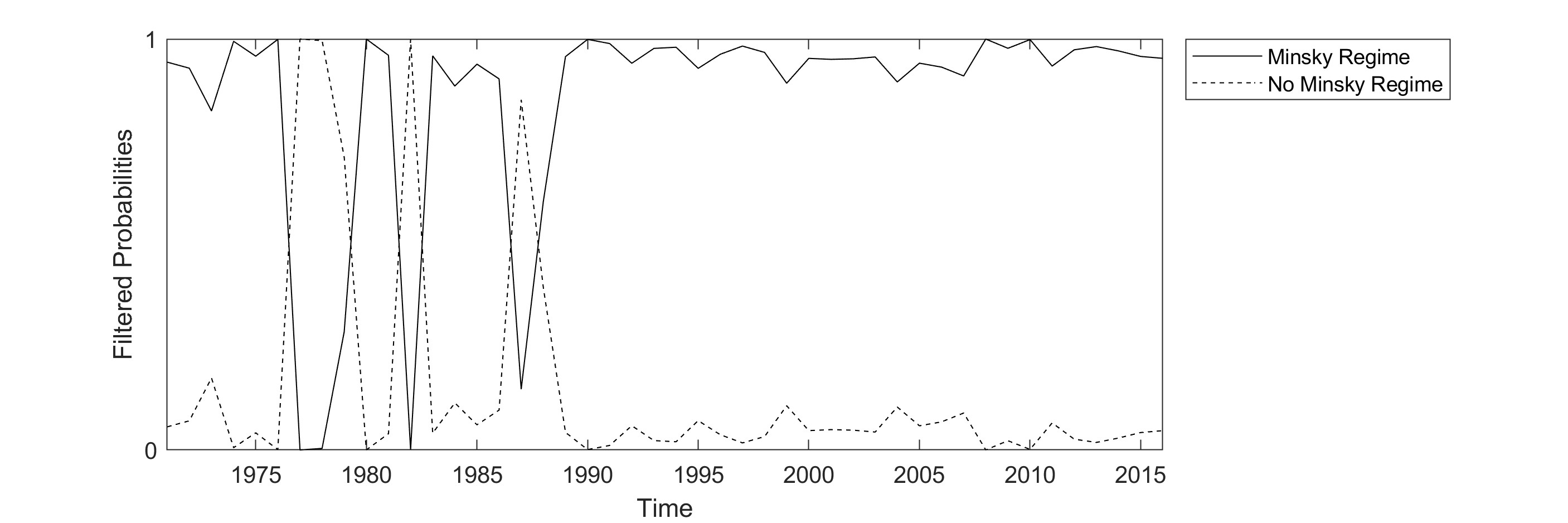}
    \caption{Filtered probability of the two regimes in the UK.}\label{filtered_Interest2}
\end{figure}

\begin{figure}[htp]
    \centering
    \includegraphics[width=0.8\linewidth]{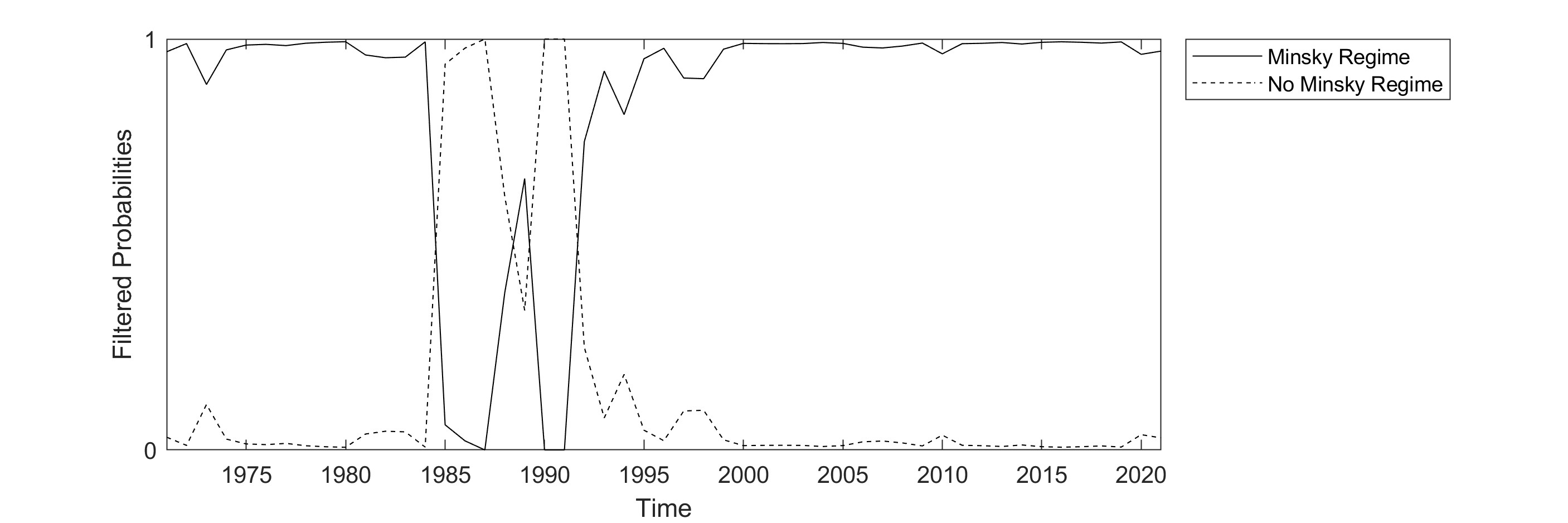}
    \caption{Filtered probability of the two regimes in Australia.}\label{filtered_Interest6}
\end{figure}

For the UK (Fig. \ref{filtered_Interest2}), prior to 1985, the system alternates between the two regimes, with brief intervals where the ``Minsky Regime" (solid line) is dominant. However, after 1985, the probability of being in a ``Minsky Regime" sharply increases, and this dynamics persists through the end of the sample period, with the "No Minsky" regime rarely appearing. These dynamics are very similar to those presented for Australia (Fig. \ref{filtered_Interest6}).

\begin{figure}[htp]
    \centering
    \includegraphics[width=0.8\linewidth]{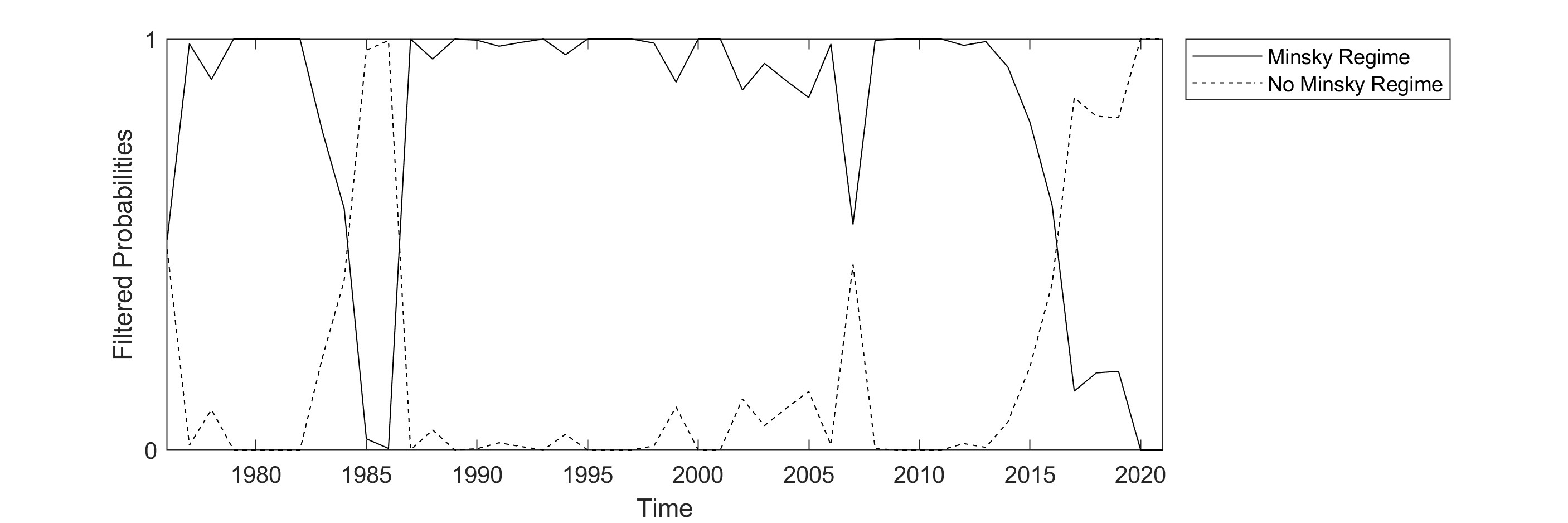}
    \caption{Filtered probability of the two regimes in France.}\label{filtered_Interest3}
\end{figure}

\begin{figure}[htp]
    \centering
    \includegraphics[width=0.8\linewidth]{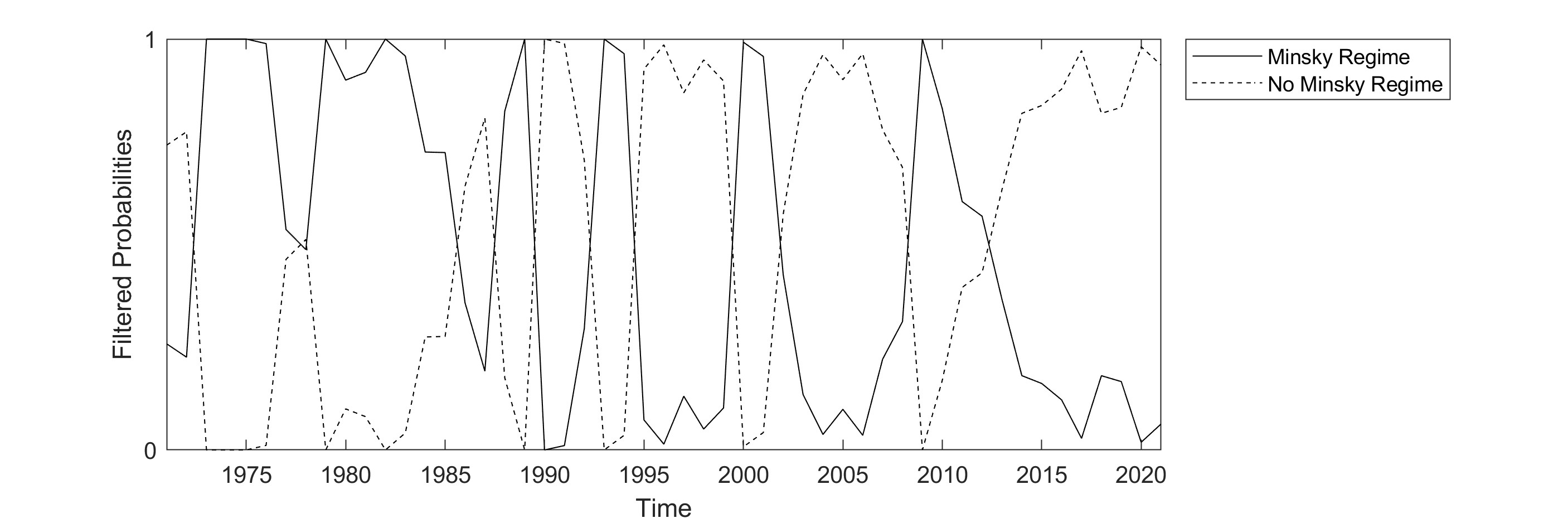}
    \caption{Filtered probability of the two regimes in Germany.}\label{filtered_Interest4}
\end{figure}

\begin{figure}[htp]
    \centering
    \includegraphics[width=0.8\linewidth]{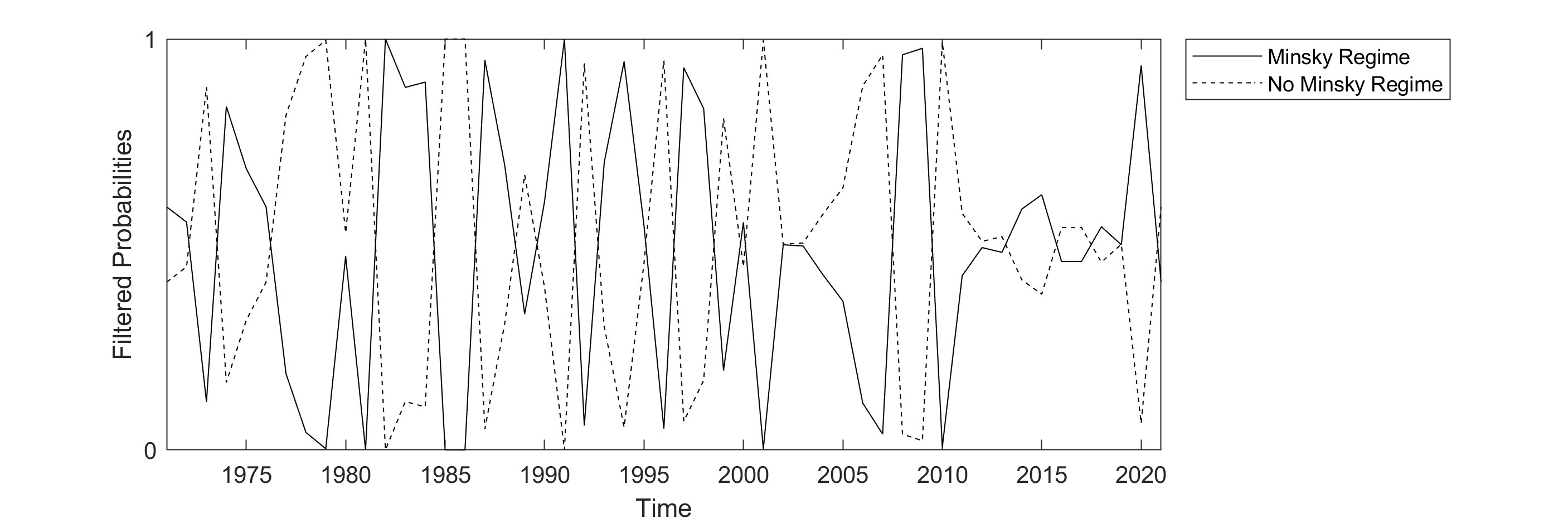}
    \caption{Filtered probability of the two regimes in Canada.}\label{filtered_Interest5}
\end{figure}

French and Germany (Fig. \ref{filtered_Interest3} and \ref{filtered_Interest4}, respectively) present similarities and differences. For both countries, from 70s to mid 80s, ``Minsky Regime" seems to dominate in term of percentage compared to the other regime. Then for France, except for the dominance of ``No Minsky Regime" in the 1985, ``Minsky Regime" becomes predominant from the mid-1990s to the late 2000s, as in the UK. Germany, on the other hand, exhibits more frequent transitions between the two regimes across the entire sample period, with no prolonged dominance of either regime. A very similar results  can be noticed  for Canada (Fig \ref{filtered_Interest5}).  These transitions highlight that neither economy maintained prolonged periods in either regime during this earlier time frame. Finally, for both France and Germany, there is a tendency toward a 'No Minsky Regime' after the 2010, a common dynamic influenced by the actions of the European Central Bank during the post-2007/2008 European crisis.

Overall, the presence of Minskyan cycles  emphasizes the crucial role that central banks play in influencing economic cycles through their interest rate policies, as shifts in interest rates can either amplify or dampen the real economy. In other words, in economies where real-financial interactions are strong, central banks’ manipulation of interest rates can trigger endogenous economic fluctuations, consistent with Minsky's financial instability hypothesis.

\begin{table}[ht]
\begin{threeparttable}
\centering
\caption{Estimation Results for GDP/STIR}\label{GDP/STIR}
\medskip

\begin{tabular}{cccc}

\hline\hline\\[-1.8ex]       
GDP/STIR & Regime 1 & Regime 2 & Transition Matrix \\
\hline\hline\\[-1.8ex] 
USA & \begin{tabular}[c]{@{}cc@{}} $0.4801^{***}$& $-1.1992^{***}$\\ (0.1293) & (0.2213) \\ $0.4484^{***}$& $0.4364^{***}$ \\ (0.0670) & (0.1147) \end{tabular} & \begin{tabular}[c]{@{}cc@{}} $0.6403^{***}$ &  \\ (0.0593) &  \\  & $0.3446^{***}$  \\  & (0.0957) \end{tabular} & \begin{tabular}[c]{@{}cc@{}}  &  \\0.537 & 0.463  \\ 0.584 & 0.416 \\ & \end{tabular} \\
\hline
& & & \\ 
\hline
UK & \begin{tabular}[c]{@{}cc@{}} $0.9328^{***}$& $-0.6669^{***}$\\ (0.0928) & (0.1251) \\ $0.3125^{***}$& $0.3440^{***}$ \\ (0.0737) & (0.0993) \end{tabular} & \begin{tabular}[c]{@{}cc@{}} $1.0135^{***}$ &  \\ (0.0681) &  \\  & $-0.3127^{*}$  \\  & (0.1684) \end{tabular} & \begin{tabular}[c]{@{}cc@{}}  &  \\ 0.898 & 0.102   \\ 0.597  & 0.403  \\ & \end{tabular} \\
\hline
& & & \\ 
\hline
France & \begin{tabular}[c]{@{}cc@{}} $0.9308^{***}$& $-0.6493^{***}$\\ (0.1196) & (0.1093) \\ $0.3892^{***}$& $0.0983^{***}$ \\ (0.1090) & (0.0996) \end{tabular} & \begin{tabular}[c]{@{}cc@{}} $0.1421^{}$ &  \\ (0.1289) &  \\  & $1.0866^{***}$  \\  & (0.0284) \end{tabular} & \begin{tabular}[c]{@{}cc@{}}  &  \\ 0.936 & 0.064  \\ 0.130  & 0.870    \\ & \end{tabular} \\
\hline
& & & \\ 
\hline              
Germany & \begin{tabular}[c]{@{}cc@{}} $0.6854^{***}$& $-0.5688^{***}$\\ (0.0967) & (0.0841) \\ $0.5028^{***}$& $0.33433^{***}$ \\ (0.1481) & (0.1288) \end{tabular} & \begin{tabular}[c]{@{}cc@{}} $0.6955^{***}$ &  \\ (0.0969) &  \\  & $0.7211^{***}$  \\  & (0.0463) \end{tabular} & \begin{tabular}[c]{@{}cc@{}}  &  \\ 0.760 & 0.240  \\ 0.222   & 0.778  \\ & \end{tabular} \\
\hline                
& & & \\ 
\hline
Canada & \begin{tabular}[c]{@{}cc@{}} $0.9885^{***}$& $-1.037^{***}$\\ (0.0905) & (0.1217) \\ $0.6089^{**}$& $0.0075^{}$ \\ (0.0570) & (0.0767) \end{tabular} & \begin{tabular}[c]{@{}cc@{}} $0.5392^{***}$ &  \\ (0.0784) &  \\  & $0.6474^{***}$  \\  & (0.1135) \end{tabular} & \begin{tabular}[c]{@{}cc@{}}  &  \\ 0.484 & 0.516   \\ 0.501  & 0.499 \\ & \end{tabular} \\
\hline      
& & & \\ 
\hline
Australia & \begin{tabular}[c]{@{}cc@{}} $0.5199^{}$& $-0.2810^{***}$\\ (0.1288) & (0.1104) \\ $0.4907^{***}$& $0.3367^{***}$ \\ (0.1091) & (0.0935) \end{tabular} & \begin{tabular}[c]{@{}cc@{}} $0.6730^{*}$ &  \\ (0.123) &  \\  & $0.1695^{***}$  \\  & (0.1515) \end{tabular} & \begin{tabular}[c]{@{}cc@{}}  &  \\ 0.970 & 0.030 \\ 0.164  & 0.836    \\ & \end{tabular} \\
\hline    
& & & \\ 
\hline
\end{tabular}
\footnotesize
 \begin{tablenotes}

            \item $^{***}$, $^{**}$, $^{*}$ are significance level at 1\%, 5\% and 10\%.
            \item Standard errors are in parenthesis.
            \item In regime one, regime two and the transition matrix, the reported values follow the positions of the parameters in section 2.1.
            
        \end{tablenotes}
    \end{threeparttable}
\label{tab:my_label}
\end{table}

\clearpage
\bigskip
\bigskip
\bigskip

\bigskip
\bigskip
\bigskip

\clearpage

\section{Conclusions}

This paper explored Minsky's financial instability hypothesis by examining the interaction mechanisms behind real-financial cycles, with a specific focus on the roles of household debt, corporate debt, and interest rates. The aim was to determine whether Minsky's hypothesis can be captured within the context of nonlinear endogenous regime changes, thereby offering a deeper understanding of the financial instability hypothesis and its implications across different countries.


To implement our analysis, we built on the framework of \citet{stockhammer2019short} and extended it by introducing a bivariate Markov-switching vector autoregressive model. This extension enabled us to capture local dynamic interaction in a context of nonlinear regime changes. Through the nonlinear filtering technique, we traced the occurrence of regime changes and observed the manifestation of Minskyan endogenous real-financial cycles during specific years between 1970 and 2020. The relationship between household debt and GDP appears comparatively weak, with significant local effects of real-financial interaction identified only in the United States and the United Kingdom.  In contrast, corporate debt exhibits more robust and widespread patterns of local interaction, observed across all countries in the sample except Australia. When short-term interest rates are incorporated into the analysis, the evidence of such interaction mechanisms becomes consistent across all countries, highlighting the central role of monetary conditions in mediating real-financial linkages.

Our study highlights the importance of considering nonlinear regime changes in the analysis of Minsky endogenous cycles,  providing a more comprehensive view of real-financial fluctuations and their underlying mechanisms. At the same time, the paper contributes to the broader literature on financial instability and macroeconomic fluctuations, providing valuable insights that could inform future economic policies aimed at mitigating financial instability.

\bibliographystyle{chicago}\bibliography{NonLinearMinsky}

\appendix

\section*{Appendix A}

Table \ref{DiagnosticCheck} shows the results of the unit root test when corporate debt, household debt and interest rate are included in the system with GDP for all the countries considered. The Dicke-Fuller test is used to test the null hypothesis that a unit root is present in the cyclical component of the series. cValue and StatValues represent the critical value and test statistics, respectively.
\bigskip
\bigskip

\begin{table}[ht]
\centering
\caption{Unit Root Test.}\label{DiagnosticCheck}
\medskip

\begin{tabular}{ccccc}

\hline\\[-1.8ex]       
USA & GDP/InterestRate  & GDP/HouseDebt & GDP/CorporateDebt  &  \\
\hline\\[-1.8ex] 

\begin{tabular}[c]{@{}c@{}} cValue \\ StatValues \end{tabular} & 
\begin{tabular}[c]{@{}c@{}} -1.94 \\ (-3.3341/-3.4015) \end{tabular} & 
\begin{tabular}[c]{@{}c@{}} -1.94 \\ (-4.4224/-2.4317) \end{tabular} & 
\begin{tabular}[c]{@{}c@{}} -1.94 \\ (-3.3341/-2.1046) \end{tabular} & \\

\hline\\[-1.8ex] 
UK  &   &  &  &  \\
\hline\\[-1.8ex] 

\begin{tabular}[c]{@{}c@{}} cValue \\ StatValues \end{tabular} & 
\begin{tabular}[c]{@{}c@{}} -1.94 \\ (-2.6198/-4.3916)  \end{tabular} & 
\begin{tabular}[c]{@{}c@{}} -1.94 \\ (-5.1540/-2.5585)  \end{tabular} & 
\begin{tabular}[c]{@{}c@{}} -1.94 \\ (-3.3733/-2.4019) \end{tabular} \\

\hline\\[-1.8ex] 
France  &   &  &  &  \\
\hline\\[-1.8ex] 

\begin{tabular}[c]{@{}c@{}} cValue \\ StatValues \end{tabular} & 
\begin{tabular}[c]{@{}c@{}} -1.94 \\ (-3.2219/-3.7994)  \end{tabular} & 
\begin{tabular}[c]{@{}c@{}} -1.94 \\ (-3.1545/-2.2614)  \end{tabular} & 
\begin{tabular}[c]{@{}c@{}} -1.94 \\ (-3.1535/-2.2069)  \end{tabular}  \\

\hline\\[-1.8ex] 
Germany &  &  &  &  \\
\hline\\[-1.8ex] 

\begin{tabular}[c]{@{}c@{}} cValue \\ StatValues \end{tabular} & 
\begin{tabular}[c]{@{}c@{}} -1.94 \\ (-3.6747/-3.7094)  \end{tabular} & 
\begin{tabular}[c]{@{}c@{}} -1.94 \\ (-4.7684/-2.3155)  \end{tabular} & 
\begin{tabular}[c]{@{}c@{}} -1.94 \\ (-3.6062/-2.4593)  \end{tabular}  \\

\hline\\[-1.8ex] 
Canada &  &  &  &  \\
\hline\\[-1.8ex] 

\begin{tabular}[c]{@{}c@{}} cValue \\ StatValues \end{tabular} & 
\begin{tabular}[c]{@{}c@{}} -1.94 \\ (-3.4071/-4.2600)  \end{tabular} & 
\begin{tabular}[c]{@{}c@{}} -1.94 \\ (-5.0253/-2.1799)  \end{tabular} & 
\begin{tabular}[c]{@{}c@{}} -1.94 \\ (-5.0253/-3.9932)  \end{tabular}  \\

\hline\\[-1.8ex] 
Australia &  &  &  &  \\
\hline\\[-1.8ex] 

\begin{tabular}[c]{@{}c@{}} cValue \\ StatValues \end{tabular} & 
\begin{tabular}[c]{@{}c@{}} -1.94 \\ (-4.3303/-3.3673)  \end{tabular} & 
\begin{tabular}[c]{@{}c@{}} -1.94 \\ (-3.3481/-3.1034)  \end{tabular} & 
\begin{tabular}[c]{@{}c@{}} -1.94 \\ (-4.5847/-2.6934)  \end{tabular} \\
\hline\hline
\end{tabular}
\end{table}

\clearpage
\section*{Appendix B}

Tables \ref{autocorrelationGDP/NFCD}, \ref{autocorrelationGDP/HD}, and \ref{autocorrelationGDP/STIR} show the results of the Ljung-Box Q-test for autocorrelation in residuals. The test is used to test the null hypothesis of no serial correlation. cValue and StatValues stand for the critical value and test statistics, respectively.

\begin{table}[ht]
\centering
\caption{Diagnostic check for residual autocorrelation (GDP/NFCD)}\label{autocorrelationGDP/NFCD}
\medskip
\begin{tabular}{ccccc}

\hline\\[-1.8ex]       
USA & $\varepsilon_t$  & $\varphi_t$  & $\delta_t$  & $\rho_t$ \\
\hline\\[-1.8ex] 

\begin{tabular}[c]{@{}c@{}} cValue \\ StatValue \end{tabular} & 
\begin{tabular}[c]{@{}c@{}} 6.63 \\ 4.39 \end{tabular} & 
\begin{tabular}[c]{@{}c@{}} 6.63 \\ 6.10 \end{tabular} & 
\begin{tabular}[c]{@{}c@{}} 6.63 \\ 0.29 \end{tabular} & 
\begin{tabular}[c]{@{}c@{}} 6.63 \\ 2.83 \end{tabular} \\

\hline\\[-1.8ex] 
UK  &   &  &  &  \\
\hline\\[-1.8ex] 

\begin{tabular}[c]{@{}c@{}} cValue \\ StatValue \end{tabular} & 
\begin{tabular}[c]{@{}c@{}} 6.63 \\ 0.03 \end{tabular} & 
\begin{tabular}[c]{@{}c@{}} 6.63 \\ 1.94 \end{tabular} & 
\begin{tabular}[c]{@{}c@{}} 6.63 \\ 0.46 \end{tabular} & 
\begin{tabular}[c]{@{}c@{}} 6.63\\ 0.47 \end{tabular} \\

\hline\\[-1.8ex] 
France  &   &  &  &  \\
\hline\\[-1.8ex] 

\begin{tabular}[c]{@{}c@{}} cValue \\ StatValue \end{tabular} & 
\begin{tabular}[c]{@{}c@{}} 6.63 \\ 0.09 \end{tabular} & 
\begin{tabular}[c]{@{}c@{}} 6.63 \\ 3.00 \end{tabular} & 
\begin{tabular}[c]{@{}c@{}} 6.63 \\ 2.20
\end{tabular} & 
\begin{tabular}[c]{@{}c@{}} 6.63 \\ 2.14 \end{tabular} \\

\hline\\[-1.8ex] 
Germany &  &  &  &  \\
\hline\\[-1.8ex] 

\begin{tabular}[c]{@{}c@{}} cValue \\ StatValue \end{tabular} & 
\begin{tabular}[c]{@{}c@{}} 6.63 \\3.18 \end{tabular} & 
\begin{tabular}[c]{@{}c@{}} 6.63 \\ 3.74 \end{tabular} & 
\begin{tabular}[c]{@{}c@{}} 6.63 \\ 0.68  \end{tabular} & 
\begin{tabular}[c]{@{}c@{}} 6.63 \\ 5.95 \end{tabular} \\

\hline\\[-1.8ex] 
Canada &  &  &  &  \\
\hline\\[-1.8ex] 

\begin{tabular}[c]{@{}c@{}} cValue \\ StatValue \end{tabular} & 
\begin{tabular}[c]{@{}c@{}} 6.63 \\ 2.18  \end{tabular} & 
\begin{tabular}[c]{@{}c@{}} 6.63 \\ 2.64
\end{tabular} & 
\begin{tabular}[c]{@{}c@{}} 6.63 \\ 1.13 \end{tabular} & 
\begin{tabular}[c]{@{}c@{}} 6.63 \\ 1.54 \end{tabular} \\

\hline\\[-1.8ex] 
Australia &  &  &  &  \\
\hline\\[-1.8ex] 

\begin{tabular}[c]{@{}c@{}} cValue \\ StatValue \end{tabular} & 
\begin{tabular}[c]{@{}c@{}} 6.63 \\ 4.30 \end{tabular} & 
\begin{tabular}[c]{@{}c@{}} 6.63 \\ 1.64 \end{tabular} & 
\begin{tabular}[c]{@{}c@{}} 6.63 \\ 0.12 \end{tabular} & 
\begin{tabular}[c]{@{}c@{}} 6.63 \\ 0.85 \end{tabular} \\
\hline\hline
\end{tabular}
\end{table}

\begin{table}[ht]
\centering
\caption{Diagnostic check for residual autocorrelation (GDP/HD)}\label{autocorrelationGDP/HD}
\medskip
\begin{tabular}{ccccc}

\hline\\[-1.8ex]       
USA & $\varepsilon_t$  & $\varphi_t$  & $\delta_t$  & $\rho_t$ \\
\hline\\[-1.8ex] 

\begin{tabular}[c]{@{}c@{}} cValue \\ StatValue \end{tabular} & 
\begin{tabular}[c]{@{}c@{}} 6.63 \\ 2.72 \end{tabular} & 
\begin{tabular}[c]{@{}c@{}} 6.63 \\ 0.88 \end{tabular} & 
\begin{tabular}[c]{@{}c@{}} 6.63 \\ 1.74 \end{tabular} & 
\begin{tabular}[c]{@{}c@{}} 6.63 \\ 5.17 \end{tabular} \\

\hline\\[-1.8ex] 
UK  &   &  &  &  \\
\hline\\[-1.8ex] 

\begin{tabular}[c]{@{}c@{}} cValue \\ StatValue \end{tabular} & 
\begin{tabular}[c]{@{}c@{}}6.63 \\ 7.75
\end{tabular} & 
\begin{tabular}[c]{@{}c@{}} 6.63 \\ 0.39 \end{tabular} & 
\begin{tabular}[c]{@{}c@{}}6.63 \\ 0.75 \end{tabular} & 
\begin{tabular}[c]{@{}c@{}} 6.63 \\ 5.25 \end{tabular} \\

\hline\\[-1.8ex] 
France  &   &  &  &  \\
\hline\\[-1.8ex] 

\begin{tabular}[c]{@{}c@{}} cValue \\ StatValue \end{tabular} & 
\begin{tabular}[c]{@{}c@{}} 6.63 \\ 0.34 \end{tabular} & 
\begin{tabular}[c]{@{}c@{}} 6.63 \\ 1.58 \end{tabular} & 
\begin{tabular}[c]{@{}c@{}} 6.63 \\ 0.85 \end{tabular} & 
\begin{tabular}[c]{@{}c@{}} 6.63 \\ 0.60 \end{tabular} \\

\hline\\[-1.8ex] 
Germany &  &  &  &  \\
\hline\\[-1.8ex] 

\begin{tabular}[c]{@{}c@{}} cValue \\ StatValue \end{tabular} & 
\begin{tabular}[c]{@{}c@{}} 6.63 \\ 1.72
\end{tabular} & 
\begin{tabular}[c]{@{}c@{}} 6.63\\ 2.37 
\end{tabular} & 
\begin{tabular}[c]{@{}c@{}} 6.63 \\ 0.86 \end{tabular} & 
\begin{tabular}[c]{@{}c@{}} 6.63 \\ 2.65 \end{tabular} \\

\hline\\[-1.8ex] 
Canada &  &  &  &  \\
\hline\\[-1.8ex] 

\begin{tabular}[c]{@{}c@{}} cValue \\ StatValue \end{tabular} & 
\begin{tabular}[c]{@{}c@{}}6.63 \\ 5.14 \end{tabular} & 
\begin{tabular}[c]{@{}c@{}}6.63 \\ 5.82 \end{tabular} & 
\begin{tabular}[c]{@{}c@{}}6.63 \\ 3.04  \end{tabular} & 
\begin{tabular}[c]{@{}c@{}}6.63 \\ 18  \end{tabular} \\

\hline\\[-1.8ex] 
Australia &  &  &  &  \\
\hline\\[-1.8ex] 

\begin{tabular}[c]{@{}c@{}} cValue \\ StatValue \end{tabular} & 
\begin{tabular}[c]{@{}c@{}}6.63 \\ 2.00\end{tabular} & 
\begin{tabular}[c]{@{}c@{}}6.63 \\ 3.57 \end{tabular} & 
\begin{tabular}[c]{@{}c@{}}6.63 \\ 0.76 \end{tabular} & 
\begin{tabular}[c]{@{}c@{}}6.63 \\ 6.43 \end{tabular} \\
\hline\hline
\end{tabular}
\end{table}

\begin{table}[ht]
\centering
\caption{Diagnostic check for residual autocorrelation (GDP/STIR)}\label{autocorrelationGDP/STIR}
\medskip
\begin{tabular}{ccccc}

\hline\\[-1.8ex]       
USA & $\varepsilon_t$  & $\varphi_t$  & $\delta_t$  & $\rho_t$ \\
\hline\\[-1.8ex] 

\begin{tabular}[c]{@{}c@{}} cValue \\ StatValue \end{tabular} & 
\begin{tabular}[c]{@{}c@{}} 6.63 \\ 3.18 \end{tabular} & 
\begin{tabular}[c]{@{}c@{}} 6.63 \\ 0.47 \end{tabular} & 
\begin{tabular}[c]{@{}c@{}} 6.63 \\ 1.92 \end{tabular} & 
\begin{tabular}[c]{@{}c@{}} 6.63 \\ 3.74 \end{tabular} \\

\hline\\[-1.8ex] 
UK  &   &  &  &  \\
\hline\\[-1.8ex] 

\begin{tabular}[c]{@{}c@{}} cValue \\ StatValue \end{tabular} & 
\begin{tabular}[c]{@{}c@{}} 6.63 \\ 0.76 \end{tabular} & 
\begin{tabular}[c]{@{}c@{}} 6.63 \\ 5.71 \end{tabular} & 
\begin{tabular}[c]{@{}c@{}} 6.63 \\ 1.00 \end{tabular} & 
\begin{tabular}[c]{@{}c@{}} 6.63 \\ 2.30 \end{tabular} \\

\hline\\[-1.8ex] 
France  &   &  &  &  \\
\hline\\[-1.8ex] 

\begin{tabular}[c]{@{}c@{}} cValue \\ StatValue \end{tabular} & 
\begin{tabular}[c]{@{}c@{}} 6.63 \\ 3.41 \end{tabular} & 
\begin{tabular}[c]{@{}c@{}} 6.63 \\ 0.72 \end{tabular} & 
\begin{tabular}[c]{@{}c@{}} 6.63 \\ 1.20 \end{tabular} & 
\begin{tabular}[c]{@{}c@{}} 6.63 \\ 5.61 \end{tabular} \\

\hline\\[-1.8ex] 
Germany &  &  &  &  \\
\hline\\[-1.8ex] 

\begin{tabular}[c]{@{}c@{}} cValue \\ StatValue \end{tabular} & 
\begin{tabular}[c]{@{}c@{}} 6.63 \\ 0.66 \end{tabular} & 
\begin{tabular}[c]{@{}c@{}} 6.63 \\ 0.05 \end{tabular} & 
\begin{tabular}[c]{@{}c@{}} 6.63 \\ 3.84 \end{tabular} & 
\begin{tabular}[c]{@{}c@{}} 6.63 \\ 3.28 \end{tabular} \\

\hline\\[-1.8ex] 
Canada &  &  &  &  \\
\hline\\[-1.8ex] 

\begin{tabular}[c]{@{}c@{}} cValue \\ StatValue \end{tabular} & 
\begin{tabular}[c]{@{}c@{}} 6.63 \\ 3.33 \end{tabular} & 
\begin{tabular}[c]{@{}c@{}} 6.63 \\ 1.04 \end{tabular} & 
\begin{tabular}[c]{@{}c@{}} 6.63 \\ 0.03 \end{tabular} & 
\begin{tabular}[c]{@{}c@{}} 6.63 \\ 0.19 \end{tabular} \\

\hline\\[-1.8ex] 
Australia &  &  &  &  \\
\hline\\[-1.8ex] 

\begin{tabular}[c]{@{}c@{}} cValue \\ StatValue \end{tabular} & 
\begin{tabular}[c]{@{}c@{}} 6.63 \\ 0.87 \end{tabular} & 
\begin{tabular}[c]{@{}c@{}} 6.63 \\ 7.35 \end{tabular} & 
\begin{tabular}[c]{@{}c@{}} 6.63 \\ 2.42 \end{tabular} & 
\begin{tabular}[c]{@{}c@{}} 6.63 \\ 8.54 \end{tabular} \\
\hline\hline
\end{tabular}
\end{table}

\clearpage

\section*{Appendix C}

In this appendix, we present the estimation results of 100 Monte Carlo simulations when GDP is included into the system with corporate debt (table \ref{MC_GDP/NFCD}), household debt (table \ref{MC_GDP/HD}) and interest rate (table \ref{MC_GDP/STIR}). 

\begin{table}[ht]
\scalebox{0.8}{
\begin{threeparttable}
\centering
\caption{Monte Carlo Estimation Results for GDP/NFCD}\label{MC_GDP/NFCD}
\medskip

\begin{tabular}{cccc}

\hline\hline\\[-1.8ex]       
GDP/NFCD & Regime 1 & Regime 2 & Transition Matrix \\
\hline\hline\\[-1.8ex] 
USA & \begin{tabular}[c]{@{}cc@{}} $0.7746$& $-0.0905$\\ (0.7115 , 0.8378) & (-0.1183 , - 0.0627) \\ $1.2018$& $0.5540$ \\ (1.0684 , 1.3351) & (0.4989 , 0.5896) \end{tabular} & \begin{tabular}[c]{@{}cc@{}} $0.4335$ &  \\ (0.3811 , 0.4859) &  \\  & $0.9385$  \\  & (0.9077 , 0.9693) \end{tabular} & \begin{tabular}[c]{@{}cc@{}}  &  \\ 0.6636 & 0.3364  \\ 0.3254  & 0.6746  \\ & \end{tabular} \\
\hline
& & & \\ 
\hline
UK & \begin{tabular}[c]{@{}cc@{}} $0.6617$& $-0.0796$\\ (0.58897 , 0.7336) & (-0.1024 , - 0.0569) \\ $1.5373$& $0.6516$ \\ (1.4123 , 1.6623) & (0.4989 , 0.5896) \end{tabular} & \begin{tabular}[c]{@{}cc@{}} $0.3298$ &  \\ (0.6268 , 0.6821) &  \\  & $0.6340$  \\  & (0.2429 , 0.4147) \end{tabular} & \begin{tabular}[c]{@{}cc@{}}  &  \\ 0.7389 & 0.2611  \\ 0.4289  & 0.5711  \\ & \end{tabular} \\
\hline
& & & \\ 
\hline
France & \begin{tabular}[c]{@{}cc@{}} $0.9672$& $-0.1423$\\ (0.9176 , 0.0168) & (-0.161 , - 0.1232) \\ $1.7109$& $0.7204$ \\ (1.6421 , 1.7797) & (0.6895 , 0.7514) \end{tabular} & \begin{tabular}[c]{@{}cc@{}} $0.3771$ &  \\ (0.3371 , 0.4171) &  \\  & $0.7862$  \\  & (0.7520 , 0.8205) \end{tabular} & \begin{tabular}[c]{@{}cc@{}}  &  \\ 0.6141 & 0.3859 \\ 0.4306  & 0.5694  \\ & \end{tabular} \\
\hline
& & & \\ 
\hline              
Germany & \begin{tabular}[c]{@{}cc@{}} $0.6090$& $-0.1793$\\ (0.5706 , 0.6473) & (-0.2019 , - 0.1566) \\ $1.7135$& $0.7391$ \\ (1.6171 , 1.8100) & (0.6983 , 0.7799) \end{tabular} & \begin{tabular}[c]{@{}cc@{}} $0.5029$ &  \\ (0.4380 , 0.5678) &  \\  & $0.6947$  \\  & (0.6077 , 0.7818) \end{tabular} & \begin{tabular}[c]{@{}cc@{}}  &  \\ 0.7230 & 0.2770  \\ 0.4825  & 0.5175  \\ & \end{tabular} \\
\hline                
& & & \\ 
\hline
Canada & \begin{tabular}[c]{@{}cc@{}} $0.1038$& $-0.3034$\\ (0.0602 , 0.1475) & (-0.3204 , - 0.2863) \\ $1.5540$& $0.5953$ \\ (1.4291 , 1.6788) & (0.5598 , 0.6318) \end{tabular} & \begin{tabular}[c]{@{}cc@{}} $0.5875$ &  \\ (0.5440 , 0.6311) &  \\  & $0.4808$  \\  & (0.4339 , 0.5277) \end{tabular} & \begin{tabular}[c]{@{}cc@{}}  &  \\ 0.5844 & 0.4156  \\ 0.3379  & 0.6221 \\ & \end{tabular} \\
\hline      
& & & \\ 
\hline
Australia & \begin{tabular}[c]{@{}cc@{}} $0.8004$& $0.0086$\\ (0.73627 , 0.8647) & (-0.0056 ,  0.0228) \\ $3.4188$& $0.6797$ \\ (3.2253 , 3.6124) & (0.6430 , 0.7164) \end{tabular} & \begin{tabular}[c]{@{}cc@{}} $0.01408$ &  \\ (-0.0448 , 0.0727) &  \\  & $0.7277$  \\  & (0.7025 , 0.7529) \end{tabular} & \begin{tabular}[c]{@{}cc@{}}  &  \\ 0.5091 & 0.4909  \\ 0.5877  & 0.4123  \\ & \end{tabular} \\
\hline    
& & & \\ 
\hline
\end{tabular}

\footnotesize
 \begin{tablenotes}           
            \item Confidence interval at 95\% level in parenthesis.
            \item In regime one, regime two and the transition matrix, the reported values follow the positions of the parameters in section 2.1.             
        \end{tablenotes}
    \end{threeparttable}}

\end{table}

\begin{table}[ht]
\scalebox{0.8}{
\begin{threeparttable}
\centering
\caption{Monte Carlo Estimation Results for GDP/HD}\label{MC_GDP/HD}
\medskip

\begin{tabular}{cccc}

\hline\hline\\[-1.8ex]       
GDP/HD & Regime 1 & Regime 2 & Transition Matrix \\
\hline\hline\\[-1.8ex] 
USA & \begin{tabular}[c]{@{}cc@{}} $0.6185$& $-0.1194$\\ (0.5317 , 0.7052) & (-0.1686 , - 0.0702) \\ $0.5783$& $0.6550$ \\ (0.4665 , 0.6900) & (0.6034 , 0.7066) \end{tabular} & \begin{tabular}[c]{@{}cc@{}} $0.2689$ &  \\ (0.2132 , 0.3245) &  \\  & $0.7985$  \\  & (0.7591 , 0.8379) \end{tabular} & \begin{tabular}[c]{@{}cc@{}}  &  \\ 0.6314 & 0.3686  \\ 0.4172  & 0.5828  \\ & \end{tabular} \\
\hline
& & & \\ 
\hline
UK & \begin{tabular}[c]{@{}cc@{}} $0.6627$& $-0.1500$\\ (0.6085 , 0.7169) & (-0.1868 , - 0.1132) \\ $0.7273$& $0.5931$ \\ (0.6694 , 0.7852) & (0.5427 , 0.6435) \end{tabular} & \begin{tabular}[c]{@{}cc@{}} $-0.0660$ &  \\ (-0.1440 , 0.0081) &  \\  & $0.3031$  \\  & (0.2221 , 0.3841) \end{tabular} & \begin{tabular}[c]{@{}cc@{}}  &  \\ 0.7231 & 0.2769  \\ 0.3855  & 0.6145  \\ & \end{tabular} \\
\hline
& & & \\ 
\hline
France & \begin{tabular}[c]{@{}cc@{}} $0.6875$& $-0.0760$\\ (0.6001 , 0.7741) & (-0.1303 , - 0.0218) \\ $0.2445$& $0.66804$ \\ (0.1893 , 0.2996) & (0.6129 , 0.7231) \end{tabular} & \begin{tabular}[c]{@{}cc@{}} $0.6933$ &  \\ (0.5975 , 0.7890) &  \\  & $0.7911$  \\  & (0.7507 , 0.8476) \end{tabular} & \begin{tabular}[c]{@{}cc@{}}  &  \\ 0.3571 & 0.6429 \\ 0.6501  & 0.3490  \\ & \end{tabular} \\
\hline
& & & \\ 
\hline              
Germany & \begin{tabular}[c]{@{}cc@{}} $0.7515$& $-0.2867$\\ (0.6483 , 0.8547) & (-0.3399 , - 0.2336) \\ $-0.0054$& $0.7258$ \\ (-0.0580 , 0.0472) & (0.6852 , 0.7665) \end{tabular} & \begin{tabular}[c]{@{}cc@{}} $0.2884$ &  \\ (0.1846 , 0.3922) &  \\  & $0.8365$  \\  & (0.8006 , 0.8724) \end{tabular} & \begin{tabular}[c]{@{}cc@{}}  &  \\ 0.3392  & 0.6608  \\ 0.6906  & 0.0394  \\ & \end{tabular} \\
\hline                
& & & \\ 
\hline
Canada & \begin{tabular}[c]{@{}cc@{}} $0.3086$& $-0.2108$\\ (0.2225 , 0.3948) & (-0.2666 , - 0.1549) \\ $0.1838$& $0.6921$ \\ (0.0638 , 0.3038) & (0.6385 , 0.7456) \end{tabular} & \begin{tabular}[c]{@{}cc@{}} $0.2160$ &  \\ (0.1448 , 0.2872) &  \\  & $0.8235$  \\  & (0.47784 , 0.8686) \end{tabular} & \begin{tabular}[c]{@{}cc@{}}  &  \\ 0.6273 & 0.3727  \\ 0.3972  & 0.6028  \\ & \end{tabular} \\
\hline      
& & & \\ 
\hline
Australia & \begin{tabular}[c]{@{}cc@{}} $0.5005$& $0.0410$\\ (0.4368 , 0.5643) & (0.0158 ,  0.0662) \\ $1.2545$& $0.4855$ \\ (1.1073 , 1.4020) & (0.4343 , 0.5367) \end{tabular} & \begin{tabular}[c]{@{}cc@{}} $0.4789$ &  \\ (0.4242 , 0.5336) &  \\  & $0.5149$  \\  & (0.4536 , 0.5762) \end{tabular} & \begin{tabular}[c]{@{}cc@{}}  &  \\ 0.6604 & 0.3396  \\ 0.4245  & 0.5755  \\ & \end{tabular} \\
\hline    
& & & \\ 
\hline
\end{tabular}
\footnotesize
 \begin{tablenotes}
 \item Confidence interval at 95\% level in parenthesis.
            \item In regime one, regime two and the transition matrix, the reported values follow the positions of the parameters in section 2.1.             
        \end{tablenotes}
    \end{threeparttable}}

\end{table}

\begin{table}[ht]
\scalebox{0.8}{
\begin{threeparttable}
\centering
\caption{Monte Carlo Estimation Results for GDP/STIR}\label{MC_GDP/STIR}
\medskip

\begin{tabular}{cccc}

\hline\hline\\[-1.8ex]       
GDP/STIR & Regime 1 & Regime 2 & Transition Matrix \\
\hline\hline\\[-1.8ex] 
USA & \begin{tabular}[c]{@{}cc@{}} $0.8474$& $-0.7936$\\ (0.8116 , 0.8832) & (-0.8551 , - 0.7322) \\ $0.5970$& $0.3398$ \\ (0.5618 , 0.6322) & (0.2979 , 0.3818) \end{tabular} & \begin{tabular}[c]{@{}cc@{}} $0.6537$ &  \\ (0.6104 , 0.6969) &  \\  & $0.6894$  \\  & (0.6534 , 0.7255) \end{tabular} & \begin{tabular}[c]{@{}cc@{}}  &  \\ 0.7026 & 0.2974  \\ 0.3184  & 0.6816  \\ & \end{tabular} \\
\hline
& & & \\ 
\hline
UK & \begin{tabular}[c]{@{}cc@{}} $0.6627$& $-0.1500$\\ (0.6085 , 0.7169) & (-0.1868 , - 0.1132) \\ $0.7273$& $0.5931$ \\ (0.6694 , 0.7852) & (0.5427 , 0.6435) \end{tabular} & \begin{tabular}[c]{@{}cc@{}} $-0.0660$ &  \\ (-0.1440 , 0.0081) &  \\  & $0.3031$  \\  & (0.2221 , 0.3841) \end{tabular} & \begin{tabular}[c]{@{}cc@{}}  &  \\ 0.7231 & 0.2769  \\ 0.3855  & 0.6145  \\ & \end{tabular} \\
\hline
& & & \\ 
\hline
France & \begin{tabular}[c]{@{}cc@{}} $0.9016$& $-0.6910$\\ (0.8544 , 0.9487) & (-0.7265 , - 0.6554) \\ $0.4446$& $0.1032$ \\ (0.4083 , 0.4909) & (0.0477 , 0.1587) \end{tabular} & \begin{tabular}[c]{@{}cc@{}} $0.2570$ &  \\ (0.1829 , 0.3311) &  \\  & $0.8353$  \\  & (0.7505 , 0.9201) \end{tabular} & \begin{tabular}[c]{@{}cc@{}}  &  \\ 0.7941 & 0.2059 \\ 0.3283  & 0.6717  \\ & \end{tabular} \\
\hline
& & & \\ 
\hline              
Germany & \begin{tabular}[c]{@{}cc@{}} $0.7640$& $-0.6091$\\ (0.7159 , 0.8122) & (-0.6483 , - 0.5700) \\ $0.5516$& $0.2605$ \\ (0.4827 , 0.6205) & (0.2059 , 0.3152) \end{tabular} & \begin{tabular}[c]{@{}cc@{}} $0.6278$ &  \\ (0.5747 , 0.6810) &  \\  & $0.7377$  \\  & (0.6926 , 0.7829) \end{tabular} & \begin{tabular}[c]{@{}cc@{}}  &  \\ 0.6828  & 0.3172  \\ 0.3280  & 0.6720  \\ & \end{tabular} \\
\hline                
& & & \\ 
\hline
Canada & \begin{tabular}[c]{@{}cc@{}} $0.9886$& $-1.0010$\\ (0.9505 , 1.0267) & (-1.0633 , - 0.9387) \\ $0.6011$& $0.0210$ \\ (0.5713 , 0.6308) & (-0.01515 , 0.0571) \end{tabular} & \begin{tabular}[c]{@{}cc@{}} $0.5005$ &  \\ (0.4568 , 0.5443) &  \\  & $0.6471$  \\  & (0.5865 , 0.7076) \end{tabular} & \begin{tabular}[c]{@{}cc@{}}  &  \\ 0.5134 & 0.4876  \\ 0.5178  & 0.4822  \\ & \end{tabular} \\
\hline      
& & & \\ 
\hline
Australia & \begin{tabular}[c]{@{}cc@{}} $0.5688$& $-0.2934$\\ (0.5152 , 0.6221) & (-0.3559 ,  -0.2309) \\ $0.4955$& $0.3129$ \\ (0.3702 , 0.6309) & (0.2528 , 0.3729) \end{tabular} & \begin{tabular}[c]{@{}cc@{}} $0.4599$ &  \\ (0.3698 , 0.5500) &  \\  & $0.3000$  \\  & (0.2010 , 0.3989) \end{tabular} & \begin{tabular}[c]{@{}cc@{}}  &  \\ 0.7176 & 0.2824  \\ 0.4882  & 0.5117  \\ & \end{tabular} \\
\hline    
& & & \\ 
\hline
\end{tabular}
\footnotesize
 \begin{tablenotes}
\item Confidence interval at 95\% level in parenthesis.
            \item In regime one, regime two and the transition matrix, the reported values follow the positions of the parameters in section 2.1.  
            
        \end{tablenotes}
    \end{threeparttable}}

\end{table}

\end{document}